\documentclass[aps,prd,onecolumn,superscriptaddress, nofootinbib, longbibliography, notitlepage]{revtex4-1}

\usepackage{graphicx}% Include figure files
\usepackage{bm}% bold math
\usepackage{amssymb}
\usepackage[utf8]{inputenc}
\usepackage{amsmath}
\usepackage{mathrsfs}
\usepackage{xcolor}
\usepackage{subcaption}
\usepackage{multirow}
\usepackage{hhline}

\RequirePackage[colorlinks,citecolor=red,urlcolor=red,linkcolor=red]{hyperref}

\def\edth{\;\raise1.0pt\hbox{$'$}\hskip-6pt\partial\;}
\def\baredth{\;\overline{\raise1.0pt\hbox{$'$}\hskip-6pt
\partial}\;}
\def\gsim{~\rlap{$>$}{\lower 1.0ex\hbox{$\sim$}}}

\newcommand{\be}{\begin{equation}}
\newcommand{\ee}{\end{equation}}
\newcommand{\bw}{\begin{widetext}}
\newcommand{\ew}{\end{widetext}}

\newcommand{\dd}{\mathrm{d}}
\newcommand{\hL}{h^{\rm L}_{A}}
\newcommand{\hNL}{h^{\rm NL}_{A}}

\newcommand{\aL}{a^{\rm L}}
\newcommand{\aNL}{a^{\rm NL}}
\newcommand{\gNL}{g_{\rm NL}}

\newcommand{\blangle}{\langle}
\newcommand{\brangle}{\rangle}

\usepackage{xcolor}
\usepackage[normalem]{ulem}

\begin{document}

%%%%%%%%%%%%%%%%%%%%
\title{Non-Gaussian timing residual skewness in pulsar timing arrays}
%%%%%%%%%%%%%%%%%%%%

\author{Reginald Christian Bernardo}
\email{reginald.christian.bernardo@aei.mpg.de}
\affiliation{Max Planck Institute for Gravitational Physics (Albert Einstein Institute), \\ Leibnitz Universit\"{a}t Hannover, D-30167 Hannover, Germany}

\author{Rituparna Mandal}
\email{drimit.ritu@gmail.com}
\affiliation{Department of Physics, National Central University, Zhongli, Taoyuan 320317, Taiwan}

\author{Kin-Wang Ng}
\email{nkw@phys.sinica.edu.tw}
\affiliation{Institute of Physics, Academia Sinica, Taipei 11529, Taiwan}
\affiliation{Institute of Astronomy and Astrophysics, Academia Sinica, Taipei 11529, Taiwan}

\date{\today}

\begin{abstract}
Motivated by searches for weak primordial non-Gaussianity in the cosmic microwave background, we investigate perturbative departures from Gaussian stochastic gravitational wave background statistics using pulsar timing arrays. We consider a quadratic ansatz that shows three-point correlations at first order in the nonlinear coupling and covariance corrections at second order. We first examine the ansatz for gravitational wave helicity amplitudes. However, the timing residual bispectrum that this produces is not manifestly invariant under local polarization basis rotations for tensor and vector modes. Then, we formulate a polarization-frame-invariant phenomenological model for detector-projected amplitudes and evaluate its one-pulsar timing residual third moment with a simplified response. Our numerical benchmarks have shown that this gives a very small Gaussian-rms-normalized skewness.
\end{abstract}
% \vskip 1cm
\maketitle

\section{Introduction}

Pulsar timing arrays (PTA) have become an invaluable tool in gravitational wave (GW) astronomy \cite{Sazhin:1978myk, Detweiler:1979wn, 1982MNRAS.199..659M, NANOGrav:2023gor, Reardon:2023gzh, EPTA:2023fyk, Xu:2023wog, InternationalPulsarTimingArray:2023mzf, Miles:2024seg, Verbiest:2024nid} through the nanohertz stochastic gravitational wave background (SGWB); complementing ground-based \cite{KAGRA:2021vkt, LIGOScientific:2021sio, ET:2025xjr} and future space-based \cite{LISA:2022kgy, Luo:2025sos} GW observations. PTA science is now in a good place to debate whether the primary source of the PTA signal is astrophysical, i.e., GWs emitted by supermassive black hole binaries (SMBHB) \cite{Sesana:2004sp, Sesana:2008mz, Kelley:2010qx, Kelley:2016gse, Kelley:2017lek, Burke-Spolaor:2018bvk, Becsy:2022pnr, Depta:2023qst, Goncharov:2024htb, Kelley:2025yud}, or cosmological (e.g., inflation, first-order phase transitions) \cite{Chen:2019xse, NANOGrav:2021flc, Xue:2021gyq, Moore:2021ibq, Bari:2023rcw, Balaji:2023ehk, NANOGrav:2023hvm, EPTA:2023xxk, Vagnozzi:2023lwo, Franciolini:2023wjm, Bai:2023cqj, Megias:2023kiy, Jiang:2023gfe, Zhang:2023nrs, Figueroa:2023zhu, Bian:2023dnv, Niu:2023bsr, Abe:2023yrw, Servant:2023mwt, Ellis:2023oxs, Bhaumik:2023wmw, Ahmed:2023pjl, Antusch:2023zjk, Aghaie:2023lan, Konoplya:2023fmh, Basilakos:2023xof, Basilakos:2023jvp, Ben-Dayan:2023lwd, Ahmadvand:2023lpp, Choudhury:2023hfm, Addazi:2023jvg, Calza:2024qxn, Domenech:2024rks, Qiu:2024sdd, Boyer:2024xvs, Tan:2024kuk, Bernardo:2025lie}. 
The Hellings \& Downs (HD) correlation provides evidence for a GW background but does not by itself distinguish its astrophysical and cosmological origins \cite{Hellings:1983fr, 1985MNRAS.217..265M, Allen:2022dzg}.
Future measurements can help distinguish source models through their spectral properties \cite{Phinney:2001di}. However, models with similar spectra and HD correlations may differ in higher-order statistics. Recent work has therefore investigated non-Gaussianity through cumulants and probability density functions (PDFs) of PTA observables \cite{Tasinato:2022xyq, Sato-Polito:2024lew, Lamb:2024gbh, Bernardo:2024uiq, Xue:2024qtx, Falxa:2025qxr, Ali-Haimoud:2026sbk}.

Different source models can produce the same mean and covariance. Higher cumulants or the full PDF can distinguish such models when their non-Gaussian signatures can be measured. A Gaussian field provides a useful reference: its mean and covariance determine all its moments through Isserlis' theorem \cite{1918Biome..12..134I}.

The cosmic microwave background (CMB) provides a motivation for this approach \cite{Planck:2019evm}. Its near Gaussian statistics are consistent with weak primordial fluctuations and higher-point correlations can distinguish scenarios with similar power spectra \cite{Planck:2018jri, Verde:2000vr, Planck:2019kim}. Tests include higher-point statistics \cite{Bernardeau:1993qu, Srednicki:1993ix, Gangui:1994wh, 1995MNRAS.274..730L, Ng:1997ez, Hui:1998ix, Komatsu:2001rj, Komatsu:2003iq}, Minkowski functionals \cite{Schmalzing:1995qn, Schmalzing:1997uc, Chingangbam:2017uqv, Bernardo:2025srg}, and peak statistics \cite{Bernardeau:1993ac, Bernardeau:2024ysi}. This motivates investigating weak departures from Gaussianity in primordial GW backgrounds.

The same motivation applies to PTAs. 
Higher-order timing residual statistics may contain information about the origin of the SGWB that is absent from its two-point function.
Following CMB and large scale structure analyses, we consider a Gaussian seed with a small quadratic correction. For this ansatz, the two-point function is unchanged at $\mathcal O(g_{\rm NL})$, while the three-point function receives an $\mathcal O(g_{\rm NL})$ contribution. This perturbative hierarchy is motivated by weak primordial non-Gaussianity, but is not exclusive to primordial sources. A finite population of SMBHBs can also generate non-Gaussian statistics \cite{Allen:2024rqk, Sato-Polito:2024lew, Xue:2024qtx}. Moreover, non-Gaussianity does not imply the existence of a nonzero skewness. A distribution symmetric about zero has a vanishing third moment. Random-phase source populations can be non-Gaussian with a vanishing ensemble bispectrum, but phase correlations can generate a bispectrum in primordial models. A nonzero third moment probes this phase-coupled component of non-Gaussianity.

In this work, we use this primordial motivation to investigate the consistency and observable consequences of a perturbative non-Gaussian ansatz \cite{Adshead:2009bz, Tsuneto:2018tif, Powell:2019kid, JimenezCruz:2025wqa}. We examine its formulation in GW helicity amplitudes and develop a detector-projected phenomenological benchmark for the timing residual third moment \cite{Bartolo:2018qqn}.

The outline of this work is as follows. In Section \ref{sec:sflrw}, we briefly review the pulsar response to GWs, expressing the result in real and harmonic spaces. The expert may skip this introductory section. In Section \ref{sec:statistics}, we present the main results of this work. We introduce statistics in the GW amplitudes (Section \ref{subsec:sgwb_statistics_primer}), and examine the polarization-frame consistency of a  non-Gaussian ansatz
formulated in terms of the GW Fourier helicity amplitude (Section \ref{subsec:no-go}). We show that the resulting timing residual
bispectrum is not manifestly polarization-frame covariant.  Then, we introduce a phenomenological ansatz in terms of the
detector-projected GW amplitudes, for which the polarization-frame
phases cancel explicitly giving a nonzero timing residual skewness (Section \ref{subsec:three_point_correlation_nonGaussian}). In Section \ref{sec:testing_gaussianity}, we estimate the nonzero timing residual skewness based on non-Gaussian detector-projected GW statistics. We discuss future work in the conclusions.

A number of appendices have been included for completeness. Appendix \ref{sec:twopoint_hd} shows the calculation of the two-point function of timing residuals. Appendix \ref{sec:generalized_projection_factors} presents a derivation of the generalized projection factors for tensor GW polarizations. Appendix \ref{sec:3pacf_Upsilon} presents analytic and numerical properties of the three-point autocorrelation function that is relevant to the evaluation of the nontrivial timing residual skewness derived in Section \ref{subsec:three_point_correlation_nonGaussian}. 

Throughout this paper, we adopt geometrized units $G = c = 1$ and the mostly-plus metric signature $(-,+,+,+)$.

\section{Pulsar response to gravitational waves}\label{sec:sflrw}

This section briefly reviews the pulsar response to passing GWs (Section \ref{subsec:pulsar_response_sachswolfe}), in real (Section \ref{subsec:real_space_formalism}) and harmonic spaces (Section \ref{subsec:harmonic_space_formalism}). Both real and harmonic space formalisms will be used to derive the results of this work.

\subsection{The response function via the Sachs-Wolfe effect}
\label{subsec:pulsar_response_sachswolfe}

Consider a superposition of GWs, with frequencies $f$, directions $\hat{k}$; and in general, with various GW polarizations, with a collective index $A$ that stands for, namely, transverse-traceless (TT) tensor modes, scalar transverse (breathing mode), scalar longitudinal, and vector modes. At a spacetime point ($\eta,\vec{x}$), the metric perturbation can be expanded in terms of its Fourier modes, each with a random, Fourier-space amplitude $h_A(\vec{k} )$;
\begin{align}
h_{i j}(\eta, \vec{x})= \sum_{A}\int_{-\infty }^{\infty} f^2 \mathrm{d} f \int_{S^{2}} \mathrm{d} \hat{k}~h_{A}(\vec{k}) e^{A}_{i j}(\hat{k})e^{-2\pi i f(\eta-v \hat{k}.\vec{x})}
\label{gw}
\end{align}
where, for convenience, we introduce the frequency-space three-vector $\vec{k}=k\hat{k}=2\pi f\hat{k}$, with $k=2\pi f$, so that $h_A(\vec{k})$ is understood as a shorthand for $h_A(f,\hat{k})$. Here, $e^{A}_{ij}(\hat{k})$ is a GW polarization-basis tensor \cite{Chamberlin:2011ev, Gair:2015hra, Qin:2018yhy, Qin:2020hfy}, and $v$ is the group velocity of the GW mode ($0<v\leq1$). For modes obeying the
standard massive dispersion relation $\omega^2=k_{\rm phys}^2+m^2$, with mass $m$, the physical spatial wavenumber is $k_{\rm phys}=\omega v=2\pi f v$. In GR, $v=1$, so that $k_{\rm phys}=k$, and the GW polarizations are $A=+,\times$.
For calculations, it is convenient to introduce the circular
polarization basis tensors, defined as
\begin{align}
e^{R}_{i j}= \frac{\left(e^{+}_{i j}+i e^{\times}_{i j} \right)}{\sqrt{2}} \, ,~~~~~~ e^{L}_{i j}= \frac{\left(e^{+}_{i j}-i e^{\times}_{i j} \right)}{\sqrt{2}}
\label{cirpol}
\end{align}
where $e^{R}_{i j}$ and $e^{L}_{i j}$ correspond to right-handed (positive-helicity) and left-handed (negative-helicity) circular polarizations, respectively. Then, the corresponding GW amplitudes appearing in \eqref{gw} will be
\begin{align}
    h_{R}=\frac{h_{+}-i h_{\times}}{\sqrt{2}} \,,\, \, \,\,\, \, \quad h_{L}=\frac{h_{+}+i h_{\times}}{\sqrt{2}}\,.
\label{eq:hRL_def}
\end{align}
We emphasize that the transverse-traceless metric
perturbation $h_{ij}$ given by \eqref{gw} represents the physical
GW field and must be independent of the arbitrary choice of polarization basis. For each propagation direction
$\hat{k}$, a local rotation of the polarization basis by an angle
$\alpha(\hat{k})$ transforms the circular polarization tensor and the
corresponding helicity amplitude as \cite{AnilKumar:2023hza}
\begin{align}
e^A_{ij}(\hat{k})
\rightarrow
e^{i s_A\alpha(\hat{k})}e^A_{ij}(\hat{k}),
\qquad \qquad
h_A(\vec{k}) \rightarrow
e^{-i s_A\alpha(\hat{k})}h_A(\vec{k}),
\end{align}
where $s_A$ denotes the corresponding spin weight. Hence,
$h_A(\vec{k})e^A_{ij}(\hat{k})$, and therefore $h_{ij}$, is invariant
under polarization-basis rotations. The same phase cancellation occurs
in the
two-point function; in particular, the auto-power
$\langle h_A h_A^*\rangle$ is polarization-frame independent.
In Sec.~\ref{subsec:no-go}, we use this requirement to examine whether
the proposed non-Gaussian ansatz yields a polarization-frame-independent
timing residual bispectrum.

Returning to \eqref{gw}, the statistics of the random variable $h_{A}(\vec{k})$ can often be approximated as Gaussian and with a continuous spectral support at nanohertz frequencies. The next section will quantify the statistics, while the rest of this section will be concerned with useful expressions of the pulsar response to GWs.

When a GW interrupts a radio pulse passing between the pulsar and the Earth, the time of arrival (ToA) of the pulse signal on Earth changes. If we take the Earth as the origin of the reference frame and $\hat{e}$ as the unit vector from Earth to the pulsar, then changes in the rotational period of the pulsar can be captured as redshift fluctuations via the Sachs-Wolfe effect \cite{Sachs:1967er, Ng:2021waj}, in a direction $\hat{e}$:
\begin{align}
z(\hat{e})= \, -\frac{1}{2}\int_{\eta_{e}}^{\eta_{r}} \mathrm{d}\eta \, \hat{e}^{i} \hat{e}^{j}\frac{\partial }{\partial \eta} h_{i j}(\eta, \vec{x})
\label{redsft}
\end{align}
where the lower and upper limits of integration in the line-of-sight integral represent the point of emission and reception, respectively, of the pulse.

The main GW observable in pulsar timing is the timing residual; or ToAs after fitting with a deterministic timing model\footnote{
This means subtracting out or marginalizing over parameters related to the pulsars' spin period, period derivatives, position, parallax, as well as the Earth's motion relative to the Solar system barycenter \cite{Hazboun:2019vhv, Taylor:2021yjx}. The process turns the residuals nonstationary, and loses the instrument sensitivity at specific frequencies like the Earth's rotational period around the Solar system barycenter \cite{Pitrou:2024scp, Allen:2025waa}.
}. The timing residual is related to the redshift via an integral;
\begin{align}
r(t, \hat{e})=\int_{0}^{t}\mathrm{d}{t}' z({t}', \hat{e}) \,.
\label{tresi}
\end{align}
Here $t$ is the elapsed time from the reference epoch $t=0$; later we denote the total observing baseline by $T$. Note that the lower limit $t'=0$ fixes $r(t=0,\hat e)=0$.
We rewrite the GW redshift fluctuations \eqref{redsft} in the laboratory time ${t}'$ as
\begin{align}
z({t}', \hat{e})=-\frac{1}{2}\int_{{t}'+\eta_{e}}^{{t}'+\eta_{r}} \mathrm{d}\eta \, d^{ij}(\hat{e})\frac{\partial }{\partial \eta} h_{i j}(\eta, \vec{x}) 
\label{rflab}
\end{align}
where we utilized the shorthand $d^{ij} (\hat{e}) \equiv \hat{e}^{i} \hat{e}^{j}$; the quantity $d^{ij}$ has been referred as the `detector tensor'. Substituting \eqref{gw} into (\ref{tresi}-\ref{rflab}), we obtain
\begin{align}
r(t,\hat{e})=-\frac{1}{2} \sum_{A} \int_{0}^{t}\mathrm{d}{t}' \int_{{t}'+\eta_{e}}^{{t}'+\eta_{r}}\mathrm{d}\eta \int_{-\infty }^{\infty}  f^2 \mathrm{d} f \int_{S^{2}} \mathrm{d} \hat{k} ~h_{A}(\vec{k}) \, d^{ij}(\hat{e})\, e^{A}_{i j} (\hat{k}) (-2\pi i f)  e^{-2\pi i f(\eta-v \hat{k}.\vec{x})} \,,
\label{trgw}
\end{align}
where $\vec{x}=\left | \vec{x} \right | \hat{e}=({t}'+\eta_{r}-\eta)\hat{e}$ is the position vector of the pulsar at time ${t}'$.\\

The following sections present two widely used expressions that are derived starting with \eqref{trgw}.

\subsection{Real space formalism}
\label{subsec:real_space_formalism}

We use the following identities to simplify the pulsar timing residual \eqref{trgw}:
\begin{align}
\int_{0}^{t}\mathrm{d}{t}' \int_{{t}'+\eta_{e}}^{{t}'+\eta_{r}}\mathrm{d}{\eta}' \, e^{-2\pi i f {\eta}'} e^{2 \pi i f v ({t}'+\eta_{r}-{\eta}') \hat{k}\cdot\hat{e} } =\frac{\left(1-e^{-2\pi i f t}\right)}{2 \pi i f} \int_{\eta_{e}}^{\eta_{r}} \mathrm{d} \eta \, e^{-2 \pi i f \eta} e^{2 \pi i f v (\eta_{r}-\eta) \hat{k}\cdot\hat{e} }
\label{ident1}
\end{align}
and 
\begin{align}
\int_{\eta_{e}}^{\eta_{r}} \mathrm{d} \eta \, e^{-2 \pi i f \eta} e^{2 \pi i f v (\eta_{r}-\eta) \hat{k}\cdot\hat{e} }=\frac{e^{-2 \pi i f \eta_{r}}}{2 \pi  f} \left[1-e^{2 \pi i f D (1+ v \hat{k} \cdot \hat{e})}\right] \frac{i}{1+v \hat{k} \cdot \hat{e} }~
\label{ident2}
\end{align}
where $D$ is the distance between the pulsar and the Earth; in terms of the pulses' emission and reception times ($\eta_r > \eta_e$), the distance is given by $D=\eta_{r}-\eta_{e} > 0$. 
Evaluating the time integral using the above identities, we obtain the timing residual GW signal given by
\begin{align}
r(t,\hat{e})=\frac{1}{4 \pi} \sum_{A}\int_{-\infty }^{\infty} f \mathrm{d}f \,\left(1-e^{-2\pi i f t}\right) e^{-2 \pi i f \eta_{r}} \int_{S^{2}} \mathrm{d} \hat{k} \, h_{A}( \vec{k}) \, d^{ij}(\hat{e})\, e^{A}_{i j} (\hat{k}) \left[1-e^{2 \pi i f D (1+ v \hat{k}\cdot\hat{e})}\right] \frac{i}{1+v \hat{k}\cdot\hat{e} }~.
\label{trgwsmp}
\end{align}
The first term in the square brackets is the so-called the `Earth term' and the second term the `pulsar term'. Note that $|v \hat{k} \cdot \hat{e} | \leq 1$; the equality is satisfied in GR ($v=1$) when the pulsar and the GW are in aligned ($\hat{k}\cdot\hat{e}=\pm1$).
Defining 
\begin{equation}
\label{eq:antenna_functions}
    F^A\left(\hat{k}, \hat{e}_a, v\right) = \dfrac{d^{ij}(\hat{e}_a) \, e_{ij}^A \left( \hat{k} \right)}{2\left( 1 + v \hat{k} \cdot \hat{e}_{{a}} \right)}
\end{equation}
and
\begin{equation}
\label{eq:Uadef}
    U\left(f D_a, \hat{k} \cdot \hat{e}_a, v\right) = 1 - e^{2 \pi i f D_a \left( 1 + v \hat{k} \cdot \hat{e}_{{a}} \right)} \,,
\end{equation}
then the timing residual induced by GWs can be written as
\begin{align}
\label{eq:timing_residual_gw_signal_rsf}
r(t,\hat{e}_a)=\frac{i}{2 \pi} \sum_{A}\int_{-\infty }^{\infty} f \mathrm{d}f \left(1-e^{-2\pi i f t}\right) e^{-2 \pi i f \eta_{r}} \int_{S^{2}} \mathrm{d} \hat{k} \, h_{A}( \vec{k}) F^A\left(\hat{k}, \hat{e}_a, v\right) U\left(f D_a, \hat{k} \cdot \hat{e}_a, v\right)
\end{align}
for a pulsar, say, labeled $a$, in the direction specified by the unit vector $\hat{e}_a$. The factors $F^A\left(\hat{k}, \hat{e}_a, v\right)$ are the so-called `antenna pattern functions' \cite{Romano:2023zhb} for a GW polarization $A$ \cite{Chamberlin:2011ev, Qin:2020hfy, NANOGrav:2021ini, Bernardo:2023zna}. We say that the signal written as \eqref{eq:timing_residual_gw_signal_rsf}, as an explicit sum over source frequencies and directions, is expressed in `real space'.

For nanohertz GW frequencies, $f = {\cal O}\left( {\rm yr}^{-1} \right) = {\cal O}\left( 31.8 \, {\rm nHz} \right)$ and for Galactic millisecond pulsars located at distances $D = {\cal O}\left( 1 \right)$ ${\rm kpc}$, the contribution of the pulsar term can be quantified by the dimensionless parameter $fD/c$ (bringing back the speed of light in vacuum). Evaluating this quantity yields
\begin{equation}
\frac{fD}{c} \simeq 103 
\left( \frac{f}{1\,{\rm nHz}} \right)
\left( \frac{D}{1\,{\rm kpc}} \right)
\simeq 3275 
\left( \frac{f}{1\,{\rm yr}^{-1}} \right)
\left( \frac{D}{1\,{\rm kpc}} \right),
\label{eq:phase_factor}
\end{equation}
where we have used $1\,{\rm kpc} = 3.086\times10^{19}\,{\rm m}$ and 
$c = 3\times10^{8}\,{\rm m/s}$. Since $fD/c \gg 1$ for typical PTA frequencies and pulsar distances,
the pulsar term oscillates rapidly across the PTA band.
Pulsar distances are also not precisely measured for most millisecond pulsars in PTAs that the phase of the pulsar term, controlled by the ratio $D/\lambda_{\rm gw}$, is unresolved compared to the GW wavelengths of interest.
Whether its products can be neglected depends on the statistic being evaluated. Unlike the Earth term, the pulsar term is also uncorrelated across pulsars down to the diffraction limit, which manifests as small scale power \cite{Ng:2021waj, 2026PhRvD.114b2003T}. For this reason, the pulsar term is often dropped in calculations or treated as a noise term in the analysis.
For us, we retain only the Earth term in the later benchmark and do not infer the accuracy of this choice for a three-point auto-statistic from the size of the phase alone.

\subsection{Harmonic space formalism}
\label{subsec:harmonic_space_formalism}

A useful way of viewing astrophysical signals is in terms of spherical harmonics, $Y_{lm}(\hat{e})$ \cite{Gair:2014rwa}. For a pulsar timing signal, we can write down
\begin{equation}
\label{eq:harmonic_sum_general}
    r\left(t, \hat{e}\right) = \sum_{lm} a_{lm}(t) Y_{lm} \left( \hat{e} \right) \,.
\end{equation}
This sum over spherical harmonics is the `harmonic space' expression for the pulsar response to GWs. Since the spherical harmonics form a complete and orthogonal basis on the two-sphere, we are guaranteed the unique set of harmonic space coefficients $a_{lm}(t)$'s; given by
\begin{equation}
    a_{lm}(t) = \int \mathrm{d} \hat{e}' \, Y_{lm}^*(\hat{e}') r(t, \hat{e}') \,.
\end{equation}

To find out what the $a_{lm}(t)$'s are for a GW signal, we return to \eqref{trgw}. We utilize the plane wave expansion in terms of Bessel functions,
\begin{equation}
    e^{i \vec{k}\cdot\vec{r}} = 4\pi \sum_{LM} i^L j_L(kr) Y_{LM}^*\left(\hat{k}\right) Y_{LM}(\hat{r}) \,, 
    \label{spherical_expans}
\end{equation}
where $j_l(x)$ is the spherical Bessel function of the first kind.
Then, the pulsar timing GW signal can be expressed as
\begin{equation}
\begin{split}
    r\left(t, \hat{e}\right) = 2\pi \sum_{LM} \sum_{A}  \int_0^t \mathrm{d}t' \int_{t' + \eta_e}^{t' + \eta_r} {\rm d} \eta
    & \int_{-\infty}^{\infty} f^2 {\rm d}f \, \left(2\pi i f\right) e^{-2\pi i f \eta} \,~ i^L j_L\left(2\pi fv \left( t' + \eta_r - \eta \right) \right)  \\
    & \,\, \times \int_{\mathbb{S}^2} {\rm d} \hat{k} \, h_A (\vec{k}) d^{ij}(\hat{e}) \, e_{ij}^A\left(\hat{k}\right) Y_{LM}^*\left(\hat{k}\right) Y_{LM}\left(\hat{e}\right) \,.
\end{split}
\label{harmonic_r}
\end{equation}
This is already in the desired form \eqref{eq:harmonic_sum_general}. We can do better. Consider the following identities\footnote{Their proof lies in changing the dummy variables, $\eta=\eta'-t'$, $x=2\pi f v(\eta_r-\eta)$, and recognizing $D = \eta_r-\eta_e$.}:
\begin{equation}
    \int_0^t {\rm d}t' \int_{t' + \eta_e}^{t' + \eta_r} {\rm d}\eta' e^{-2\pi i f \eta'} W \left( t' + \eta_r - \eta' \right) = \left( \dfrac{1 - e^{-2\pi i f t}}{2\pi i f} \right) \int_{\eta_e}^{\eta_r} {\rm d}\eta \, e^{-2\pi i f \eta} \, W\left( \eta_r - \eta \right)
\end{equation}
and
\begin{equation}
    \int_{\eta_e}^{\eta_r} {\rm d} \eta \, e^{-2\pi i f \eta} \, j_L \left( 2\pi f v \left(\eta_r - \eta\right) \right) = \left( \dfrac{e^{-2\pi i f \eta_r}}{2\pi f v} \right) \int_0^{2\pi f D v} {\rm d} x \, e^{i x/v} j_L\left(x\right) \,.
\end{equation}
Then, we arrive at our desired expression:
\begin{equation}
\begin{split}
    r\left(t,\hat{e}\right) = \ & \sum_{LM} \sum_A \int_{-\infty}^\infty f \, {\rm d}f \left( 1 - e^{-2\pi i f t} \right) e^{-2\pi i f \eta_r} \\
    & \,\,\,\, \times \int_{\mathbb{S}^2}  {\rm d}\hat{k} \, h_A\left( \vec{k}\right) \left[ d^{ij}(\hat{e}) \, e_{ij}^A\left( \hat{k} \right) \right] Y^*_{LM}\left(\hat{k}\right) \left( i^L \int_0^{2\pi f Dv} \dfrac{{\rm d}x}{v} e^{ix/v}  j_L\left(x\right) \right) Y_{LM}\left(\hat{e}\right) \,.
\end{split}
\label{eq:timing_residual_gw_signal_hsf}
\end{equation}
In this form, the pulsar timing residual harmonic space coefficients $a_{lm}(t)$'s in \eqref{eq:harmonic_sum_general} can be read as
\begin{equation}
\label{eq:timing_residual_harmonic_coefficients}
\begin{split}
    a_{lm}(t) = \ & \sum_A \int_{-\infty}^\infty f \, {\rm d}f \left( 1 - e^{-2\pi i f t} \right) e^{-2\pi i f \eta_r} \int_{\mathbb{S}^2}  {\rm d}\hat{k} \, h_A(\vec{k}) J^{A}_{lm}\left(fD, v, \hat{k} \right)
\end{split}
\end{equation}
where
\begin{equation}
 J^{A}_{lm}\left(fD, v, \hat{k} \right)= \sum_{LM} Y^*_{LM}\left(\hat{k}\right)  \left( i^L \int_0^{2\pi f Dv} \dfrac{{\rm d}x}{v} e^{ix/v} j_L\left(x\right) \right) \, \int_{\mathbb{S}^2} \mathrm{d} \hat{e} \, \left[ d^{ij}(\hat{e}) \, e_{ij}^A\left( \hat{k} \right) \right] Y_{LM}(\hat{e}) \,  Y_{lm}^*(\hat{e})\,.
 \label{redshift_fluc}
\end{equation}

\section{GW statistics in the two- and three-point correlation functions}
\label{sec:statistics}

We derive our main results in this section, starting with a brief review of GW statistics (Section \ref{subsec:sgwb_statistics_primer}). Then, utilizing the harmonic space expression \eqref{eq:harmonic_sum_general} of the pulsar timing residual, we examine the polarization-frame consistency of the helicity-space non-Gaussian ansatz and show that the resulting timing residual bispectrum is not manifestly invariant under local rotations of the
polarization basis (Section \ref{subsec:no-go}). We subsequently introduce a
phenomenological non-Gaussian ansatz for the detector-projected GW
amplitude, for which the polarization-frame phases cancel explicitly (Section \ref{subsec:three_point_correlation_nonGaussian}).

\subsection{SGWB statistics: Gaussian vs perturbative non-Gaussian}
\label{subsec:sgwb_statistics_primer}

When looking at a superposition \eqref{gw}, the case (that we want to confront) is that the number of sources is very large and the astrophysical distributions unknown. A traditional way to deal with this is to associate the GWs, $h_{ij}(\eta, \vec{x})$, or its Fourier amplitudes, $h_A(\vec{k})$, to a random variable that is specified by its moments, or in full, a PDF. 

Working with $h_A(\vec{k})$ is particularly convenient because the Fourier modes of a Gaussian field decouple. This implies that the statistical properties of $h_{ij}(\eta,\vec{x})$ can be understood in terms of the behavior of individual Fourier modes, providing a straightforward route to computing correlation functions and power spectra.

Before specifying the statistics, we recall the reality condition of
the metric perturbation. As defined after \eqref{gw},
$\vec{k}=2\pi f\hat{k}$, where the frequency $f$ takes both positive
and negative values. Under complex conjugation, the frequency changes
as $f\to -f$ while the direction $\hat{k}$ is kept fixed, so that
$\vec{k}\to-\vec{k}$. Here and below, $-\vec{k}$ in the reality
relation therefore denotes $2\pi(-f)\hat{k}$ with the same $\hat{k}$.
Then, the reality condition
$h_{ij}(\eta,\vec{x})=h_{ij}^{*}(\eta,\vec{x})$ implies
\begin{align}
\sum_A h_A(-\vec{k})e^A_{ij}(\hat{k})
=
\sum_A h_A^*(\vec{k})e^{A*}_{ij}(\hat{k}) \,.
\end{align}
With the real linear-polarization basis,
$e^{+*}_{ij}=e^+_{ij}$ and $e^{\times *}_{ij}=e^\times_{ij}$, we have
\begin{align}
h_+(-\vec{k})=h_+^*(\vec{k}) \,,
\qquad
h_\times(-\vec{k})=h_\times^*(\vec{k}) \,.
\end{align}
With the circular-polarization basis,
$e^{R*}_{ij}=e^L_{ij}$ and $e^{L*}_{ij}=e^R_{ij}$, we have
\begin{align}
h_R(-\vec{k})=h_L^*(\vec{k}) \,,
\qquad
h_L(-\vec{k})=h_R^*(\vec{k}) \,.
\end{align}
We introduce the notation $\bar A$ to denote the polarization label
associated with complex conjugation of the corresponding polarization
tensor,
$e^{A*}_{ij}=e^{\bar A}_{ij}$ \,.
The reality condition implies
\begin{align}
h_A(-\vec{k})=h_{\bar A}^{*}(\vec{k}) \,.
\label{reality_relation}
\end{align}
where, in our signed-frequency convention,
$-\vec{k}=2\pi(-f)\hat{k}$ with $\hat{k}$ kept fixed.
For the linear-polarization basis, complex conjugation leaves the
polarization label unchanged. For the circular-polarization
basis it interchanges the two helicities. Explicitly,
\begin{align}
\bar +=+ \,,\qquad
\bar\times=\times \,,\qquad
\bar R=L \,,\qquad
\bar L=R \,.
\end{align}

The angular integrations in this formalism run over the full sphere of propagation directions. This does not require a source at every sky position. Discrete sources can be represented by angular delta functions under the same integral. Nor does the general covariance below assume isotropy. Statistical isotropy is a symmetry of ensemble statistics, rather than uniformity of each sky realization. An isotropic, unpolarized ensemble is adopted later for the quantitative benchmark in Section~\ref{subsec:three_point_correlation_nonGaussian}.

For a Gaussian field, the mean and covariance specify its statistics:
\begin{align}
\label{eq:mean_general}    \langle h_A(\vec{k}) \rangle = \, & 0 \\
\label{eq:covariance_general_Star}  
\left\langle
h_A(\vec{k})h_{A'}^*(\vec{k}')
\right\rangle
={}&
(2\pi)^3\delta^{(3)}(\vec{k}-\vec{k}')
P_{AA'}(\vec{k}) \,,
\end{align}
where ${P}_{A A'} ( \vec{k} )$ is the auto- or cross-power spectra for $A=A'$ and $A\neq A'$, respectively, for GW polarizations $A$ and $A'$. The higher odd-moments vanish and even-moments can be derived using Isserlis' theorem \cite{1918Biome..12..134I}.
Using the reality relation
$h_A(-\vec{k})=h_{\bar A}^*(\vec{k})$, the same covariance can
equivalently be written in the all-unstarred form as
\begin{align}
\label{eq:covariance_general}
\left\langle
h_A(\vec{k})h_{A'}(\vec{k}')
\right\rangle
=
(2\pi)^3\delta^{(3)}(\vec{k}+\vec{k}')
P_{A\bar A'}(\vec{k}) \,.
\end{align}

For a non-Gaussian field, higher moments are not in general determined by the covariance. We consider a quadratic perturbation that preserves the zero mean and the Gaussian covariance at first order in the coupling, while producing a leading-order third moment. Isserlis' theorem remains valid for the Gaussian seed and will be used to evaluate the perturbative corrections.
We decompose the Fourier modes of gravitational perturbations into a linear (Gaussian) part and a nonlinear contribution,
\begin{align}
h_{A}(\vec{k}) \equiv h^{\rm L}_{A}(\vec{k}) + h^{\rm NL}_{A}(\vec{k}) \,,
\label{eq:hNL_sum}
\end{align}
where the  term $h^{\rm L}(\vec{k})$ satisfies Gaussian statistics described in the previous paragraph, e.g., (\ref{eq:mean_general}-\ref{eq:covariance_general}) with $h_{A}(\hat{k})$'s replaced with $h^{\rm L}_{A}(\hat{k})$'s.
Inspired by CMB analyses \cite{Komatsu:2001rj, Komatsu:2003iq}, the term $h^{\rm NL}(\vec{k})$ is taken to be
\begin{align}
\label{eq:hnl_general}
h_{A}^{\rm NL}(\vec{k})=g_{\rm NL} \left( \int \frac{ \mathrm{d}^{3}\vec{p}}{(2\pi)^3}\, h^{\rm L}_{A} ( \vec{k} + \vec{p} ) h^{\rm L *}_{A}\left(\vec{p} \right) - (2\pi)^3 \, \delta^{(3)}(\vec{k}) \langle | \hL(\eta, \vec{x}) |^2 \rangle \right) \,,
\end{align}
where $h_A^{\rm L}(\eta, \vec{x})$ denotes the real-space field associated
with the Fourier amplitude $h_A^{\rm L}(\vec{k})$. The equal-time, equal-point
variance in \eqref{eq:hnl_general} can be written as
\begin{equation}
    \langle |\hL\left(\eta, \vec{x}\right)|^2 \rangle = \frac{1}{(2\pi)^3}\int \dd^3 \vec{p} ~\, {\cal P}_A \left(f_{p}\hat{p}\right) \,,
    \label{eq:variance}
\end{equation}
where $\vec{p}=p\hat{p}=2\pi f_{p}\hat{p}$\, and $P_{AA}(\vec{p}) \equiv {\cal P}_A( f_p \hat{p} )$.
It can be shown that the first and second moments are given by
\begin{align}
\label{eq:mean_hNL_sum}
   \blangle h_A (\vec{k}) \brangle = \,& 0 \, \\
\label{eq:covariance_hNL_sum}
    \langle h_{A}(\vec{k}) h_{A'} (\vec{k}') \rangle = \,& (2\pi)^3\delta^{(3)} ( \vec{k}+ \vec{k}' ) P_{A \bar A'} ( f \hat{k} ) + {\cal O}\left( \gNL^2 \right) \,.
\end{align}
The cross terms at ${\cal O}\left(\gNL\right)$ vanishes,
\begin{align}
 \blangle \hL(\vec{k}) h_{A'}^{\rm NL}(\vec{k}') \brangle = 0 \,,
 \label{eq:hLhNLave}
 \end{align}
because odd-moments of the Gaussian random variable $h^{\rm L}_A(\hat{k})$ vanishes. A non-Gaussian contribution arises in
\begin{align}
 \blangle \hNL (\vec{k}) ^2 \brangle = {\cal O} \left( \gNL^2 \right) \,.
 \label{eq:hNL2ave}
\end{align}

In PTAs, the first two moments are sufficient to show that the leading GW signal due to a SGWB is the timing residual correlation, or two-point correlation function, characterized by the HD curve \cite{Allen:2022dzg, Bernardo:2022xzl, Allen:2024uqs}; but also see \cite{Tasinato:2022xyq} for a caveat. Consequently, the statistics of the Gaussian $h_A^{\rm L}(\vec{k})$ and perturbatively non-Gaussian $h_A(\vec{k})$ cannot be distinguished with PTA two-point correlation function measurements. Since this is very well studied, we skip the details in the main text and refer the interested readers to Appendix \ref{sec:twopoint_hd}. The point is that, due to (\ref{eq:mean_hNL_sum}-\ref{eq:covariance_hNL_sum}), even with a non-Gaussian GW random field, the timing residual two-point correlation function will take the same form as that of a Gaussian field, $h_A(\vec{k})$ defined by (\ref{eq:hNL_sum}-\ref{eq:variance}), until ${\cal O}(g_{\rm NL}^2)$ terms are included. 
This motivates studying observables directly based on higher-moments, such as the skewness and kurtosis, and higher-point correlation function to test SGWB statistics with PTAs.
When the SGWB is Gaussian, all odd higher-point correlation functions vanish, while the even higher-point statistics of the GW amplitudes satisfy Isserlis’ theorem \cite{1918Biome..12..134I, Allen:2024bnk}\footnote{In particle physics, this result is known as Wick’s theorem. Historically, Wick discovered the theorem much later compared to Isserlis.}. In this case, the cosmic variance naturally arises \cite{Allen:2022dzg, Bernardo:2022xzl}.

As we alluded to in the introduction, the first two moments alone generally cannot tell random signals apart. To illustrate this point, we consider higher moments. Using (\ref{eq:hNL_sum}-\ref{eq:covariance_hNL_sum}), it can be shown that\footnote{In writing \eqref{eq:ensemble_average_g_pert}, we have used
\begin{align}
  \delta^{(3)} \left( \vec{k} +  \vec{k}' +  \vec{k}'' \right) = \frac{1}{(2\pi)^3}\delta^{(3)} \left( f\hat{k} + f' \hat{k}' + f'' \hat{k}'' \right) \,. 
  \label{eq:scaling}
\end{align}}
\begin{equation}
    \blangle \hL(\vec{k}) h_{A'}^{\rm L}(\vec{k}') h_{A''}^{\rm NL}(\vec{k}'') \brangle = \gNL \delta^{(3)} ( f\hat{k} + f' \hat{k}' + f'' \hat{k}'' ) \left( P_{A \bar A''}(f \hat{k}) P_{A' A''} ( f'\hat{k}' ) +
P_{A A''}(f\hat{k}) P_{A'\bar A''}(f'\hat{k}') \right )\,.
    \label{eq:ensemble_average_g_pert}
\end{equation}
This contributes a leading order, ${\cal O}\left( \gNL \right)$ non-Gaussian (bispectrum) correction to the ensemble statistics; i.e., the third moment is given by
\begin{equation}
\begin{split}
    \blangle h_{A}(\vec{k}) h_{A'}(\vec{k}') h_{A''}(\vec{k}'') \brangle & = \blangle \hL(\vec{k}) h_{A'}^{\rm L}(\vec{k}') h_{A''}^{\rm NL}(\vec{k}'') \brangle + \blangle \hL(\vec{k}) h_{A'}^{\rm NL}(\vec{k}') h_{A''}^{\rm L}(\vec{k}'') \brangle + \blangle h_{A}^{\rm NL}(\vec{k}) h_{A'}^{\rm L}(\vec{k}') h_{A''}^{\rm L}(\vec{k}'') \brangle \,.
\end{split}
\end{equation}
Since the three-point function of the Gaussian part of the Fourier modes of the gravitational perturbations vanishes,
the nontrivial contribution to the third moment of the random variable $h_A^{\rm L}(\vec{k})$ can be associated with non-Gaussianity:
\begin{align}
\blangle h_{A}(\vec{k}) h_{A'}(\vec{k}') h_{A''}(\vec{k}'') \brangle  = {\cal O}\left( \gNL \right) \,.
\label{eq:h3nongaussian}
\end{align}
The perturbative expansion requires the nonlinear contribution to be small relative to the Gaussian seed over the scales considered; the coupling alone does not set this condition independently of the field amplitude. The hierarchy of the two- and three-point corrections motivates the use of higher moments when the covariance alone cannot distinguish the models at leading order.

In the following section, we will work with (\ref{eq:hNL_sum}–\ref{eq:hnl_general}) and compute three-point timing residual correlation function for a PTA experiment.

\subsection{
Polarization-basis dependence of the timing residual three-point function
}
\label{subsec:no-go}

The polarization-frame issue can be seen directly in the ansatz. Under the local rotation $h_A^{\rm L}(\vec q)\rightarrow e^{-is_A\alpha(\hat q)}h_A^{\rm L}(\vec q)$, its quadratic integrand transforms as
\begin{align}
h_A^{\rm L}(\vec k+\vec p)h_A^{\rm L*}(\vec p)
\longrightarrow {}& e^{-is_A[\alpha((\vec k+\vec p) / |\vec k+\vec p| )-\alpha(\hat p)]}
h_A^{\rm L}(\vec k+\vec p)h_A^{\rm L*}(\vec p) \,.
\label{eq:quadratic_frame_phase}
\end{align}
This phase generally differs from $e^{-is_A\alpha(\hat k)}$, which is required of an output helicity amplitude at $\vec k$. A scalar-inspired convolution without a compensating spin-dependent kernel is therefore not manifestly covariant \cite{JimenezCruz:2025wqa}. 

For the present perturbative ansatz, Section \ref{subsec:sgwb_statistics_primer} has discussed that the two-point timing residual correlation for a Gaussian GW field $h_A(\vec{k})$ is indistinguishable to that of a perturbatively non-Gaussian field $h_A(\vec{k})$ satisfying (\ref{eq:hNL_sum}-\ref{eq:variance}), up to ${\cal O}(g_{\rm NL}^2)$ corrections. In symbols,
\begin{equation}
\label{eq:2pcf_ng}
    \blangle r\left(t, \hat{e}_a\right) r\left(t', \hat{e}_b\right) \brangle = \blangle r\left(t, \hat{e}_a\right) r\left(t', \hat{e}_b\right) \brangle _ {g_{\rm NL} \rightarrow 0} + {\rm O}(g_{\rm NL}^2) \,,
\end{equation}
where the first term on the right, $\blangle r\left(t, \hat{e}_a\right) r\left(t', \hat{e}_b\right) \brangle _ {g_{\rm NL} \rightarrow 0}$, is the two-point function for the corresponding Gaussian field $h_A(\vec{k})$.
The explicit results in real and harmonic space are shown in Section \ref{sec:twopoint_hd}. See \cite{Tasinato:2022xyq} for a counterpoint for a stationary non-Gaussian SGWB.

Following the hint of \eqref{eq:h3nongaussian}, we compute the three-point timing residual correlation function:
\begin{equation}
\label{eq:3pcf_def}
    \blangle r\left(t, \hat{e}_a\right) r\left(t', \hat{e}_b\right) r\left(t'', \hat{e}_c\right) \brangle \,.
\end{equation}
This describes the correlation between timing measurements of three pulsars, with distances and positions labeled by $a, b, c$; $D_a$ and $\hat{e}_a$ are the distance and sky position of pulsar $a$, $D_b$ and $\hat{e}_b$ of pulsar $b$, and $D_c$ and $\hat{e}_c$ of pulsar $c$. The measurement \eqref{eq:3pcf_def} is performed at a time $t$, $t'$ and $t''$ for pulsars $a$, $b$, and $c$.
We compute the three-point function \eqref{eq:3pcf_def} in harmonic space, utilizing \eqref{eq:timing_residual_harmonic_coefficients}. In this case, we obtain
\begin{equation}
    \blangle r\left(t, \hat{e}_a\right) r\left(t', \hat{e}_b\right) r\left(t'', \hat{e}_c\right) \brangle = \sum_{l_1 m_1} \sum_{l_2 m_2} \sum_{l_3 m_3} \blangle a_{l_{1}m_{1}} a_{l_{2}m_{2}} a_{l_{3}m_{3}} \brangle Y_{l_1 m_1}\left( \hat{e}_a \right) Y_{l_2 m_2}\left( \hat{e}_b \right) Y_{l_3 m_3}\left( \hat{e}_c \right) + {\cal O}\left( \gNL^2 \right) \,,
\end{equation}
The non-Gaussian statistics of the SGWB can be accommodated in harmonic space by writing
\begin{align}
    a_{lm}=\aL_{lm}+ \aNL_{lm}~ \,,
\end{align}
reflective of the decomposition of \eqref{eq:hNL_sum} into Gaussian and non-Gaussian pieces.
Then, the angular bispectrum \cite{Komatsu:2001rj, Komatsu:2003iq, Gair:2014rwa} relevant for PTA analyses is defined as
\begin{align}
    B^{m_{1} m_{2} m_{3}}_{l_{1} l_{2} l_{3}}&= \blangle a_{l_{1}m_{1}} a_{l_{2}m_{2}} a_{l_{3}m_{3}} \brangle \, \nonumber \\ &= \blangle \aL_{l_{1}m_{1}} \aL_{l_{2}m_{2}} \aNL_{l_{3}m_{3}} \brangle +\blangle \aL_{l_{1}m_{1}} \aNL_{l_{2}m_{2}} \aL_{l_{3}m_{3}} \brangle +\blangle \aNL_{l_{1}m_{1}} \aL_{l_{2}m_{2}} \aL_{l_{3}m_{3}} \brangle \,, 
    \label{eq:bispectrum}
\end{align}
where the leading-order nontrivial contribution at ${\cal O}(\gNL)$ arises from the three terms shown above. In what follows, we focus on the first term. Substituting \eqref{eq:timing_residual_harmonic_coefficients} into \eqref{eq:bispectrum}, and using \eqref{eq:ensemble_average_g_pert},
we obtain
 \begin{align}
\blangle \aL_{l_{1}m_{1}} \aL_{l_{2}m_{2}} \aNL_{l_{3}m_{3}} \brangle &=  g_{\rm NL} \, \sum_{A A'A''} \int f \, \dd f \, c(f,t,\eta_r) \int f' \, \dd f' \, c(f',t',\eta_r) \int f'' \,\dd f'' \, c(f'',t'',\eta_r) \int \mathrm{d}\hat{k} \int \mathrm{d}{\hat{k}}' \nonumber \\ & 
  \int \mathrm{d}{\hat{k}}'' \, \, \delta^{(3)}\left(f\hat{k}+{f}'\hat{k}'+{f}''\hat{k}'' \right) \, \left( P_{A \bar A''}(f \hat{k}) P_{A' A''} ( f'\hat{k}' ) +
P_{A A''}(f\hat{k}) P_{A'\bar A''}(f'\hat{k}') \right )\nonumber \\ & \qquad \qquad  J^{A}_{l_1m_1}\left(fD, v, \hat{k} \right)  J^{A'}_{l_2m_2}\left(f'D, v, {\hat{k}}' \right) J^{A''}_{l_3m_3}\left(f''D, v, {\hat{k}}'' \right)\,,
\label{angular_bispec_eq1}
 \end{align}
where we defined
\begin{equation}
    c(f, t, \eta_r) = \left(1-e^{-2\pi i f t}\right) e^{-2\pi i f \eta_r} \,.
    \label{eq:cftn}
\end{equation} 
To evaluate this, we use the integral representation of the Dirac delta function,
\begin{align}
\delta^{(3)}(\vec{k})=\frac{1}{(2 \pi)^3}\int_{\mathbb{R}^3} \mathrm{d}^{3}\vec{x}~ e^{i \vec{k} \cdot \vec{x}} \,,
\label{Dd}
\end{align}
and expand the exponential factors in terms of spherical harmonics using \eqref{spherical_expans}. The angular bispectrum becomes
\begin{align}
&\langle \aL_{l_{1}m_{1}} \aL_{l_{2}m_{2}} \aNL_{l_{3}m_{3}} \rangle = 8 \, g_{\rm NL} \, \sum_{A A'A''} \int f \, \dd f \, c(f,t,\eta_r) \int f' \, \dd f' \, c(f',t',\eta_r) \int f'' \,\dd f'' \, c(f'',t'',\eta_r) \int \mathrm{d}\hat{k} \int \mathrm{d}\hat{k}' \int \mathrm{d}\hat{k}''  \nonumber \\ & 
 \sum_{l_4 \, l_5 \, l_6 \atop m_4 m_5 m_6}  i^{l_{4}+l_{5}+l_{6}}  \sqrt{\frac{\left(2l_{4}+1\right)\left(2l_{5}+1\right)\left(2l_{6}+1\right)}{4\pi}}\begin{pmatrix}
l_{4} & l_{5} & l_{6}\\
0 & 0 & 0
\end{pmatrix}
 \begin{pmatrix}
l_{4} & l_{5} & l_{6}\\
m_{4} & m_{5} & m_{6}
\end{pmatrix} J_{l_{4}l_{5}l_{6}}(f,{f}',{f}'') Y_{l_{4}m_{4}}(\hat{k})Y_{l_{5}m_{5}}({\hat{k}}')  \nonumber \\ &  Y_{l_{6}m_{6}}(\hat{k}'')
 \,  \left( P_{A \bar A''}(f \hat{k}) P_{A' A''} ( f'\hat{k}' ) +
P_{A A''}(f\hat{k}) P_{A'\bar A''}(f'\hat{k}') \right )  J^{A}_{l_1m_1}\left(fD, v, \hat{k} \right)  J^{A'}_{l_2m_2}\left(f'D, v, {\hat{k}}' \right) J^{A''}_{l_3m_3}\left(f''D, v, \hat{k}'' \right)
 \label{Bispectrum}
 \end{align}
where we have defined
\begin{align}
J_{l_{1}l_{2}l_{3}}(f,{f}',{f}'') \equiv \int_{0}^{\infty} \mathrm{d}x ~x^{2} j_{l_{1}}(fx) \, j_{l_{2}}({f}'x)\, j_{l_{3}}({f}''x)
\label{jll}
\end{align}
for convenience. In deriving the above expression, we have performed the angular integration arising from $\dd^{3}x = x^{2} \, \dd x  \, \dd \hat{x}$ and used the Gaunt integral,
\begin{align}
\int \mathrm{d}\hat{x} Y_{l_{1}m_{1}}(\hat{x})Y_{l_{2}m_{2}}(\hat{x}) Y_{l_{3}m_{3}}(\hat{x})=\sqrt{\frac{(2l_{1}+1)(2l_{2}+1)(2l_{3}+1)}{4\pi}} \begin{pmatrix}
l_{1} & l_{2} & l_{3}\\
0 & 0 & 0
\end{pmatrix}
 \begin{pmatrix}
l_{1} & l_{2} & l_{3}\\
m_{1} & m_{2} & m_{3}
\end{pmatrix}~.
\label{Gaunt}
\end{align}

To proceed further, we need the generalized projection factors $J^{A}_{lm}(fD,  v, \hat{k})$ \cite{Gair:2015hra, Qin:2020hfy, Bernardo:2022vlj, Bernardo:2022rif}. Fortunately, the relevant quantities have already been derived \cite{Qin:2020hfy, Bernardo:2022vlj, Bernardo:2022rif, Liu:2022skj}, which we review in Appendix \ref{sec:generalized_projection_factors} and utilize in their final form. 
For tensor GW modes, $J^{A}_{l m}(fD,v,\hat{k})$ takes the form as
\begin{equation}
\label{eq:Jlm_tensor}
    J_{lm}^{R,L}\left(fD, v, \hat{k}\right) = - _{\mp 2}Y_{lm}^* \left( \hat{k} \right) e^{\mp 2i \alpha} \sqrt{2}\pi i^l \sqrt{ \dfrac{(l + 2)!}{(l - 2)!} } \int_0^{2\pi fDv} \dfrac{dx}{v} \ e^{ix/v} \dfrac{j_l(x)}{x^2} \,,
\end{equation}
where the upper (lower) signs correspond $R$ ($L$) polarizations, respectively.
For vector gravitational waves, the corresponding $J_{lm}\left(fD, v, \hat{k}\right)$ coefficients can be expressed as
\begin{equation}
\label{eq:Jlm_vector}
\begin{split}
    J_{(l \geq 1)m}^{VR,VL} \left( fD, v, \hat{k} \right) = \mp \, _{\mp 1} Y_{lm}^*\left(\hat{k}\right) e^{\mp i \alpha} 2\sqrt{2} \pi i^l \sqrt{l (l + 1)} \int_0^{2\pi f D v} \dfrac{dx}{v} \ e^{ix/v} \dfrac{d}{dx} \left( \dfrac{j_l(x)}{x} \right) \,.
\end{split}
\end{equation}
The angle $\alpha$ in \eqref{eq:Jlm_tensor} and \eqref{eq:Jlm_vector} represents the freedom to rotate the polarization basis around each GW propagation direction
$\hat{k}$ and may therefore depend on $\hat{k}$ \cite{AnilKumar:2023hza}. Under such a change,
the response coefficient $J^{A}_{l m}$ acquires a
spin-dependent phase, which must be cancelled by the corresponding
phase of the helicity amplitude or its correlation functions in a
physical observable.

Consider the angular integration on $\dd \hat{k}''$ in calculating the three point timing residual correlation from Eq. \eqref{Bispectrum},
\begin{align}
  \int_{\mathbb{S}^2} \dd \hat{k}'' Y_{l_{6}m_{6}}(\hat{k}'')   J^{A''}_{l_3m_3}\left(f''D, v, \hat{k}'' \right)  \,.
  \label{50}
\end{align} 
In \eqref{50}, the ordinary spherical harmonic has zero spin weight,
whereas $J^{A''}_{l_3m_3}$ carries the polarization-basis
phase. Since the helicity bispectrum in \eqref{eq:ensemble_average_g_pert}, derived from the non-Gaussian ansatz in (\ref{eq:hNL_sum}-\ref{eq:hnl_general}), does not display the
additional spin-dependent factor required to cancel this phase, the
bispectrum in \eqref{Bispectrum} is not manifestly independent of the
polarization-basis choice. 

The same issue applies to the $\hat{k}$ and $\hat{k}'$ integrations
in \eqref{Bispectrum}, where the polarization-indexed power
spectra must transform together with the corresponding response
coefficients. If these spectra are treated as scalar functions and
provide no compensating spin phases, these integrals, and hence the bispectrum, remain polarization-basis dependent. Since the
timing residual bispectrum is a physical quantity, this indicates a
lack of polarization-frame covariance in the adopted ansatz in (\ref{eq:hNL_sum}-\ref{eq:hnl_general}).

A fully polarization-frame independent harmonic-space treatment would require either a suitable generalization of the non-Gaussian ansatz (\ref{eq:hNL_sum}-\ref{eq:hnl_general}) or an equivalent spin-covariant structure in the resulting helicity bispectrum in \eqref{eq:ensemble_average_g_pert}, such that the spin phases of all three
response coefficients are compensated \cite{Bartolo:2018qqn}. Determining the required spin
factors and constructing such a polarization-frame-covariant
bispectrum is left for future work. In the following section, we
instead introduce a phenomenological ansatz directly in terms of the
detector-projected GW amplitude, for which the polarization-frame
phases cancel explicitly.

\subsection{
A detector-projected non-Gaussian timing residual skewness
}
\label{subsec:three_point_correlation_nonGaussian}

In the derivation below, timing residual skewness refers to the unnormalized third moment $\mu_3=\langle r^3\rangle$. The dimensionless skewness $\gamma_1$ and its Gaussian-rms normalization $\mathcal K$ are distinguished in \eqref{eq:normalized_skewness}.

The analysis of subsections \ref{subsec:sgwb_statistics_primer} and \ref{subsec:no-go} shows that the helicity-space ansatz in (\ref{eq:hNL_sum}-\ref{eq:hnl_general}) does not manifestly yield a
polarization-frame independent timing residual bispectrum. In this section, we
consider an alternative phenomenological construction in
which non-Gaussian statistics are introduced directly in the
detector-projected GW amplitude. 

GW detection relies on time delays induced by passing GWs and detector-projected quantities; for PTAs, we have
$d^{ij}(\hat{e}) h_{ij}(t, \vec{x})$, where the detector tensor $d^{ij}(\hat{e}) \equiv \hat{e}^i \hat{e}^j$ is built on unit vectors corresponding to pulsar directions with Earth as the origin. The relation to testing statistics is the following. For a Gaussian SGWB, the statistics of the Fourier modes $h_A(\hat{k})$ is reflected in the statistics of the time-domain waveform $h_{ij}(t, \vec{x})$, and consequently on the timing residuals $r(t, \hat{e}_a)$. However, when modelling non-Gaussianity and computing observables, one gets to choose between invoking non-Gaussian statistics in the Fourier modes, $h_A(\hat{k})$, or in terms of the polarization-resolved detector-projected amplitudes $h_{A}(\vec{k})d^{i j}(\hat{e})e^{A}_{i j}(\hat{k})$, whose sum over polarizations reconstructs the detector projection $d^{ij}(\hat{e}) h_{ij}(t, \vec{x})$. The former gives polarization-basis-dependent timing residual skewness within
the helicity-space ansatz (Section~\ref{subsec:no-go}), whereas the
latter provides an alternative phenomenological construction, as we shall see explicitly in the remainder of this section.

Define the detector-projected GW amplitude as
\begin{align}
\label{eq:htilde_def}
\tilde{h}_{A}(\vec{k},\hat{e})\equiv h_{A}(\vec{k})d^{i j}(\hat{e})e^{A}_{i j}(\hat{k})\,.  
\end{align}
Using the reality relation \eqref{reality_relation} and $e^{A*}_{ij}=e^{\bar A}_{ij}$, the detector-projected amplitudes satisfy $\tilde h_A(-\vec k,\hat e)=\tilde h_{\bar A}^*(\vec k,\hat e)$.
Under a rotation of the local polarization basis, the helicity
amplitude $h_A(\vec{k})$ and the polarization tensor
$e^{A}_{ij}(\hat{k})$ acquire opposite spin phases. Their product,
and consequently the detector-projected amplitude $\tilde{h}_{A}(\vec{k},\hat{e})$, is invariant under polarization-frame rotations. Formulating the non-Gaussian ansatz in terms of this projected quantity makes the required phase cancellation explicit.

Motivated by the principles of GW detection, we directly work with $\tilde{h}_{A}(\vec{k},\hat{e})$.
Then, analogous to (\ref{eq:hNL_sum}-\ref{eq:hnl_general}), we write down the detector-projected GW amplitude as a sum of Gaussian $\tilde{h}_{A}^{\rm L}(\vec{k},\hat{e})$ and non-Gaussian parts $\tilde{h}_{A}^{\rm NL}(\vec{k},\hat{e})$;
\begin{align}
\tilde{h}_{A}(\vec{k},\hat{e})=\tilde{h}_{A}^{\rm L}(\vec{k},\hat{e})+\tilde{h}_{A}^{\rm NL}(\vec{k},\hat{e})\,,
\label{gwfur}
\end{align}
where
\begin{align}
\tilde{h}_{A}^{\rm NL}(\vec{k},\hat{e})=g_{\rm NL} \left( \int \frac{\mathrm{d}^{3}\vec{p}}{(2\pi)^3} \, \tilde{h}_{A}^{\rm L}(\vec{k} + \vec{p},\hat{e}) \, \tilde{h}^{\rm L *}_{A}(\vec{p},\hat{e})-(2\pi)^3\delta^{(3)}(\vec{k}) \blangle | \tilde{h}^{\rm L}_{A}(\eta,\vec{x},\hat{e})|^{2}\brangle \right)~.
\label{gwnon}
\end{align}

Equation \eqref{eq:htilde_def} specifies the projection of the Gaussian seed used to construct these channels. Once \eqref{gwnon} is imposed separately for each direction $\hat e$, the nonlinear channels are a phenomenological model; we have not shown that they can all be written as projections of a single detector-independent nonlinear tensor field. Polarization-frame invariance alone does not establish that stronger property.

The construction is used as an equal-time, nonstationary phenomenological benchmark. The momentum-triangle constraint is retained without additionally imposing the signed-frequency condition $f+f'+f''=0$ required for a stationary three-point spectrum \cite{JimenezCruz:2025wqa}. The reference epoch $\eta_r$ in \eqref{eq:cftn} cannot in general be shifted without affecting the model. This issue is separate from the nonstationarity introduced by fixing $r(0,\hat e)=0$ or fitting a timing model.

The nonvanishing result that we obtain does not conflict with \cite{JimenezCruz:2025wqa}; which finds that, under their assumptions, polarization-angle averaging leads to vanishing three-point detector correlations for tensor and vector gravitational-wave modes. The construction we considered is different in two important respects. First, the nonlinear projected channels are defined directly at the detector-response level, and we have not established that they can be obtained as projections of a single detector-independent nonlinear tensor field. Second, the present benchmark is explicitly nonstationary and does not impose the stationary signed-frequency condition assumed in the usual stochastic-background treatment \cite{Tsuneto:2018tif, Powell:2019kid, JimenezCruz:2025wqa}. The nonzero third moment derived below should therefore be interpreted as a phenomenological, detector-channel-level result rather than as a generic prediction for an isotropic tensor SGWB.

The first term in \eqref{gwfur} represents a Gaussian random field, such that its statistics are fully defined by the mean and the covariance,
\begin{align}
\langle \tilde{h}_{A}^{\rm L}(\vec{k},\hat{e}) \rangle &= 0 \,, \label{mean_hL_dp} \\
\blangle \tilde{h}_{A}^{\rm L}(\vec{k},\hat{e}_{1}) \, \tilde{h}_{{A}'}^{\rm L *}(\vec{k}',\hat{e}_{2}) \brangle &= (2\pi)^3 \delta^{(3)}(\vec{k}-\vec{k}') \, \tilde{P}_{A{A}'}(f, \hat{k}, \hat{e}_1, \hat{e}_2) \,, \label{twpnt}
\end{align}
where
\begin{equation}
    \tilde{P}_{A{A}'}(f, \hat{k}, \hat{e}_1, \hat{e}_2) = P_{A{A}'}(f\hat{k}) \, d^{i j}(\hat{e}_{1})e^{A}_{i j}(\hat{k})\, d^{k l}(\hat{e}_{2})e^{{A}'*}_{k l}({\hat{k}})
\end{equation}
is the power spectrum of the detector-projected GW amplitude $\tilde{h}_{A}^{\rm L}(\vec{k},\hat{e})$, and all the higher moments can be derived by applying Isserlis theorem \cite{1918Biome..12..134I}. The second term in \eqref{gwfur} defined by \eqref{gwnon} represents a leading-order non-Gaussian component and is proportional to the coupling parameter $g_{\rm NL}$.
It can be shown that the first and second moments are
\begin{align}
\label{eq:mean_hNL_sum_dp}
   \blangle \tilde{h}_A (\vec{k}, \hat{e}) \brangle = \,& 0 \, \\
\label{eq:covariance_hNL_sum_dp}
    \langle \tilde{h}_{A}(\vec{k}, \hat{e}_1) \tilde{h}_{A'} (\vec{k}', \hat{e}_2) \rangle = \,& (2\pi)^3\delta^{(3)} ( \vec{k}+ \vec{k}' ) P_{A \bar{A'}} ( f \hat{k} ) \, d^{i j}(\hat{e}_{1})e^{A}_{i j}(\hat{k})\, d^{k l}(\hat{e}_{2})e^{{A}'}_{k l}(\hat{k'}) + {\cal O}\left( \gNL^2 \right) \,.
\end{align}
The barred polarization index appears because \eqref{eq:covariance_hNL_sum_dp} is written in the all-unstarred form. Using $\tilde h_A(-\vec k,\hat e)=\tilde h_{\bar A}^*(\vec k,\hat e)$, \eqref{eq:covariance_hNL_sum_dp} is equivalent to the starred covariance in \eqref{twpnt}.
The cross terms at ${\cal O}\left(\gNL\right)$ vanish,
\begin{align}
 \blangle \tilde{h}_A^L(\vec{k}, \hat{e}) \tilde{h}_{A'}^{\rm NL}(\vec{k}', \hat{e}) \brangle = 0 \,,
 \label{eq:hLhNLave_dp}
 \end{align}
because odd-moments of $h^{\rm L}_A(\hat{k})$ vanish. At the level of the two-point function, the leading-order non-Gaussian contribution arises in
\begin{align}
 \blangle \tilde{h}^{NL}_{A} (\vec{k}, \hat{e}) ^2 \brangle = {\cal O} \left( \gNL^2 \right) \,.
 \label{eq:hNL2ave_dp}
\end{align}
Therefore, two-point correlation functions of the detector-projected GW amplitude $\tilde{h}_{A}(\vec{k},\hat{e})$, and the observables tied to it, such as the two-point timing residual correlation and the power spectrum, are insensitive to the non-Gaussianity at leading order in $\gNL$.

The leading-order non-Gaussian contribution arises in the three-point correlation function. The three-point correlation function of the detector-projected GW amplitude $\tilde{h}_{A}(\vec{k},\hat{e})$ is given by 
\begin{align}
\blangle \tilde{h}_{A}(\vec{k},\hat{e}_{1}) \,\tilde{h}_{{A}'}(\vec{k}',\hat{e}_{2}) \, \tilde{h}_{{A}''}(\vec{k}'',\hat{e}_{3})\brangle = \,& \blangle \tilde{h}_{A}^{\rm L}(\vec{k},\hat{e}_{1})\,\tilde{h}_{{A}'}^{\rm L}(\vec{k}',\hat{e}_{2}) \, \tilde{h}_{{A}''}^{\rm NL}(\vec{k}'',\hat{e}_{3})\brangle \nonumber \\ & +  \blangle \tilde{h}_{A}^{\rm L}(\vec{k},\hat{e}_{1})\,\tilde{h}_{{A}'}^{\rm NL}(\vec{k}',\hat{e}_{2}) \, \tilde{h}_{{A}''}^{\rm L}(\vec{k}'',\hat{e}_{3})\brangle  \nonumber \\ &+ \blangle \tilde{h}_{A}^{\rm NL}(\vec{k},\hat{e}_{1})\,\tilde{h}_{{A}'}^{\rm L}(\vec{k}',\hat{e}_{2}) \,\tilde{h}_{{A}''}^{\rm L}(\vec{k}'',\hat{e}_{3}) \brangle  + \mathcal{O}(\gNL^2) \,.
\label{nvthr}
\end{align}
The contributions $\blangle\tilde{h}_{A}^{\rm L}(\vec{k},\hat{e}_{1})\, \tilde{h}_{{A}'}^{\rm L}(\vec{k}',\hat{e}_{2}) \, \tilde{h}_{{A}''}^{\rm L}(\vec{k}'',\hat{e}_{3})\brangle =0$ vanish exactly due to $h_A^{\rm L}(\hat{k}, \hat{e})$ being a zero-mean Gaussian field.
The three-point function of the detector-projected GW amplitude can be expressed as
\begin{align}
&\blangle \tilde{h}_{A}^{\rm L}(\vec{k},\hat{e}_{1}) \, \tilde{h}_{{A}'}^{\rm L}(\vec{k}',\hat{e}_{2}) \, \tilde{h}_{{A}''}^{\rm NL}(\vec{k}'',\hat{e}_{3})\brangle = \gNL \, \delta^{(3)}(f\hat{k}+{f}'\hat{k}'+{f}''\hat{k}'') \, \Big( P_{A\bar A''}(f\hat{k})\,
P_{A'A''}(f'\hat{k}') \mathcal{D}_A(\hat{k},\hat{e}_{1})\,
\mathcal{D}_{A''}(\hat{k},\hat{e}_{3})\,
\nonumber\\
&\qquad\qquad\times
\mathcal{D}_{A'}(\hat{k}',\hat{e}_{2})\,
\mathcal{D}^{*}_{A''}(\hat{k}',\hat{e}_{3})
+
P_{AA''}(f\hat{k})\,
P_{A'\bar A''}(f'\hat{k}')
\mathcal{D}_{A}(\hat{k},\hat{e}_{1})\,
\mathcal{D}^{*}_{A''}(\hat{k},\hat{e}_{3})\,
\mathcal{D}_{A'}(\hat{k}',\hat{e}_{2})\,
\mathcal{D}_{A''}(\hat{k}',\hat{e}_{3})
\Big)\,
\label{tcorf}
\end{align}
where we make use of the shorthand $\mathcal{D}_{A}(\hat{k},\hat{e})\equiv d^{ij}(\hat e)e^{A}_{ij}(\hat k)$.

For the rest of this section, we show that the PTA three-point autocorrelation function, or timing residual skewness, is non-vanishing at leading order in $\gNL$ when the non-Gaussianity is introduced in the detector-projected GW amplitude $\tilde{h}_{A}(\vec{k},\hat{e})$. Our final expression is given by \eqref{eq:mu3_full_sign_sectors} together with the definitions (\ref{finangc}-\ref{eq:beta_def}). The reader interested in observational consequences may go directly to the final result and the discussion that follows. The derivation is provided for completeness.

We adopt two simplifying assumptions for the quantitative benchmark. First, we retain only the Earth term. Although the pulsar phase in \eqref{eq:phase_factor} is large for typical PTA frequencies and distances, rapid oscillation alone does not establish that all pulsar-term products are negligible in a three-point auto-statistic. Their contribution has not been calculated here.\footnote{Coherent use of pulsar terms can improve localization and angular resolution \cite{Allen:2022dzg, Ng:2021waj, 2026PhRvD.114b2003T}.} Second, we replace the exact antenna response inside the integrals by $F^A(\hat k,\hat e,v)\sim d^{ij}(\hat e)e^A_{ij}(\hat k)/2$. This retains the detector projection but omits the PTA directional denominator. We use it as a simplified response, without assuming that the correction to the cubic statistic is only an order-one factor. For the rest of the quantitative calculation we also restrict the ensemble to an isotropic, unpolarized SGWB containing the two tensor polarizations. In the circular basis, $P_{AA'}(f\hat k)=Y_{00}(\hat k)\delta_{AA'}P_{AA}(f)$, with $A,A'=R,L$ and $P_{RR}(f)=P_{LL}(f)$. These are additional ensemble assumptions and do not identify the source as primordial.

With these approximations, the timing residual skewness or three-point autocorrelation function for a single pulsar can be written as
\begin{align}
\langle r(t,\hat{e})^{3}\rangle \sim \, & \left(\frac{i}{4\pi}\right)^{3} \sum_{A A' A''}\int f \, \dd f \, c(f,t,\eta_r) \int  f' \, \dd f'\, \, c(f',t,\eta_r) \int f'' \, \dd f'' \, c(f'',t,\eta_r) \nonumber \\ & \quad \int \mathrm{d}\hat{k} \int \mathrm{d}\hat{k}' \int \mathrm{d}\hat{k}'' \, \blangle \tilde{h}_{A} ( \vec{k},\hat{e} ) \,\tilde{h}_{A'} (\vec{k}',\hat{e} ) \tilde{h}_{{A}''} ( \vec{k}'',\hat{e} )\brangle \,,
\label{tpcor}
\end{align}
where we have utilized the definitions \eqref{eq:htilde_def} and \eqref{eq:cftn} for the detector-projected GW amplitude and the pulsar temporal response function. The pulsar label in the direction $\hat{e}$ has been suppressed, as we restrict our analysis to a single pulsar.
Utilizing (\ref{nvthr}-\ref{tpcor}), we obtain the skewness of the timing residual, up to $\mathcal{O}(\gNL)$, as
\begin{align}
\langle r(t,\hat{e})^{3}\rangle \sim \,&  g_{\rm NL} \left(\frac{i}{4\pi}\right)^{3} \sum_{A A' A''}\int f \, \dd f \, c(f,t,\eta_r) \int  f' \, \dd f'\, \, c(f',t,\eta_r) \int f'' \, \dd f'' \, c(f'',t,\eta_r) \int \mathrm{d}\hat{k} \int \mathrm{d}\hat{k}' \int \mathrm{d}\hat{k}''  \nonumber \\
&   \delta^{(3)} (f\hat{k}+{f}'\hat{k}'+{f}''\hat{k}'' ) \, \Big( P_{A\bar A''}(f\hat{k})\,
P_{A'A''}(f'\hat{k}') \mathcal{D}_A(\hat{k},\hat{e})\,
\mathcal{D}_{A''}(\hat{k},\hat{e})\,
\mathcal{D}_{A'}(\hat{k}',\hat{e})\,
\mathcal{D}^{*}_{A''}(\hat{k}',\hat{e}) \nonumber \\
& \qquad\qquad\qquad\qquad\qquad +
P_{AA''}(f\hat{k})\,
P_{A'\bar A''}(f'\hat{k}')
\mathcal{D}_{A}(\hat{k},\hat{e})\,
\mathcal{D}^{*}_{A''}(\hat{k},\hat{e})\,
\mathcal{D}_{A'}(\hat{k}',\hat{e})\,
\mathcal{D}_{A''}(\hat{k}',\hat{e})
\Big) \nonumber \\
& + \left\lbrace {f}'{\hat{k}}', {A}' \leftrightarrow  {f}'' {\hat{k}}'', {A}'' \right\rbrace + \left\lbrace f\hat{k},A \leftrightarrow  {f}'' {\hat{k}}'',{A''} \right\rbrace + \mathcal{O}(\gNL^2) ~.
\label{tpcorre}
\end{align}
The symbol $\lbrace {f}'{\hat{k}}', {A}' \leftrightarrow  {f}'' {\hat{k}}'', {A}'' \rbrace$ denotes the term obtained by interchanging the primed and double-primed variables in the preceding expression, and similarly for the other term. This accounts for the three permutations of the three-point correlation function.
The Dirac delta function is evaluated using \eqref{spherical_expans}, \eqref{Dd}, and \eqref{Gaunt}. This leads to the expression for the three-point autocorrelation function of the timing residuals in terms of spherical harmonics and Wigner 3j symbols, given by:
\begin{align}
\langle r(t,\hat{e})^{3}\rangle \sim \,
& 8 \, g_{\rm NL} \left(\frac{i}{4\pi}\right)^{3} \, \sum_{A A' A''}\int f \, \dd f \, c(f,t,\eta_r) \int  f' \, \dd f'\, \, c(f',t,\eta_r) \int f'' \, \dd f'' \, c(f'',t,\eta_r) \int \mathrm{d}\hat{k} \int \mathrm{d}{\hat{k}}' \int \mathrm{d}{\hat{k}}'' \nonumber \\
& \sum_{l_{1} \, l_{2} \, l_{3} \atop m_{1} m_{2} m_{3}} i^{l_{1}+l_{2}+l_{3}} \sqrt{\frac{\left(2l_{1}+1\right)\left(2l_{2}+1\right)\left(2l_{3}+1\right)}{4\pi}} \begin{pmatrix}
l_{1} & l_{2} & l_{3}\\
0 & 0 & 0
\end{pmatrix}
 \begin{pmatrix}
l_{1} & l_{2} & l_{3}\\
m_{1} & m_{2} & m_{3}
\end{pmatrix}  J_{l_{1}l_{2}l_{3}}(f,{f}',{f}'') \, \nonumber \\
&   Y_{l_{1}m_{1}}(\hat{k}) \,Y_{l_{2}m_{2}}({\hat{k}}') Y_{l_{3}m_{3}}({\hat{k}}'') \,
  \Big( P_{A\bar A''}(f\hat{k})\,
P_{A'A''}(f'\hat{k}') \mathcal{D}_A(\hat{k},\hat{e})\,
\mathcal{D}_{A''}(\hat{k},\hat{e})\,
\mathcal{D}_{A'}(\hat{k}',\hat{e})\,
\mathcal{D}^{*}_{A''}(\hat{k}',\hat{e})
\nonumber\\
&
\qquad\qquad\qquad\qquad\qquad\qquad\,\,+
P_{AA''}(f\hat{k})\,
P_{A'\bar A''}(f'\hat{k}') \mathcal{D}_{A}(\hat{k},\hat{e})\,
\mathcal{D}^{*}_{A''}(\hat{k},\hat{e})\,
\mathcal{D}_{A'}(\hat{k}',\hat{e})\,
\mathcal{D}_{A''}(\hat{k}',\hat{e})
\Big) \nonumber \\
& + \left\lbrace {f}'{\hat{k}}', {A}' \leftrightarrow  {f}'' {\hat{k}}'', {A}'' \right\rbrace + \left\lbrace f\hat{k},A \leftrightarrow  {f}'' {\hat{k}}'',{A''} \right\rbrace + \mathcal{O}(\gNL^2) \phantom{\dfrac{1}{1}} \,.
\label{finaltpc}
\end{align}
Recall that $J_{l_{1}l_{2}l_{3}}(f,{f}',{f}'')$ is defined in \eqref{jll}. The remaining task now is to evaluate the angular integrals in \eqref{finaltpc} to obtain our final expression for the three-point autocorrelation function as a three-dimensional frequency integral. It turns out that the angular integrals can be evaluated analytically.

Let the pulsar direction $\hat{e}$ be aligned with the polar axis, or $\hat{z}$ axis, and the GW propagate in the direction $-\hat{k}$ with wave vector $\hat{k}$. In this coordinate system, the GW polarization tensors can be expressed in terms of spherical coordinate basis vectors as
\begin{align}
e^{+}_{i j}&=\hat{m} \otimes \hat{m}- \hat{n} \otimes \hat{n} \nonumber \\
e^{\times}_{i j}&=\hat{m} \otimes \hat{n}+ \hat{n} \otimes \hat{m}
\end{align}
where $({\hat{k},\hat{m},\hat{n}})$ forms an orthonormal basis defined by $\hat{k} \cdot \hat{m} = \hat{k} \cdot \hat{n} = \hat{m} \cdot \hat{n} = 0$ and $\hat{k} \times \hat{m} = \hat{n}$, with the explicit expressions
$\hat{k} \equiv (\sin \theta \cos \phi, ~ \sin\theta \sin\phi, ~\cos\theta)$, $\hat{m}\equiv(\sin\phi,~ -\cos \phi,~ 0)$, and $\hat{n} \equiv (\cos\theta \cos\phi, ~\cos\theta \sin\phi,  ~-\sin\theta)$.
Working with the circular polarization basis and utilizing the isotropy of the SGWB, the three-point autocorrelation function of the timing residuals can be expressed as
\begin{align}
  \langle r(t,\hat{e})^{3}\rangle \sim \, & 2 \gNL \left(\frac{i}{4\pi}\right)^{3}\int f \, \dd f \, c(f,t,\eta_r) \int  f' \, \dd f'\, \, c(f',t,\eta_r) \int f'' \, \dd f'' \, c(f'',t,\eta_r) \nonumber \\
  & \quad  \left( P_{L L}(f) \,  P_{R R}({f}')+ P_{R R}(f) \,  P_{L L}({f}') \right) \, \Upsilon (f,{f}',{f}'')  + \left\lbrace {f}'\leftrightarrow  {f}'' \right\rbrace + \left\lbrace f  \leftrightarrow  {f}''  \right\rbrace \,,
\label{tpfcr}
\end{align}
where
\begin{align}
\Upsilon (f,{f}',{f}'') \equiv \, & 8 \int \mathrm{d}\hat{k} \int \mathrm{d}{\hat{k}}' \int \mathrm{d}{\hat{k}}''  \,  \sum_{l_{1} \, l_{2} \, l_{3} \atop m_{1} m_{2} m_{3}} i^{l_{1}+l_{2}+l_{3}} \sqrt{\frac{\left(2l_{1}+1\right)\left(2l_{2}+1\right)\left(2l_{3}+1\right)}{4\pi}}
\begin{pmatrix}
l_{1} & l_{2} & l_{3} \\
0 & 0 & 0
\end{pmatrix}
\nonumber \\  
& \quad 
\begin{pmatrix}
l_{1} & l_{2} & l_{3}\\
m_{1} & m_{2} & m_{3}
\end{pmatrix}
 J_{l_{1}l_{2}l_{3}}(f,{f}',{f}'') \, Y_{00}(\hat{k}) \,Y_{00}({\hat{k}}')\, Y_{l_{1}m_{1}}(\hat{k}) \,Y_{l_{2}m_{2}}({\hat{k}}') \, Y_{l_{3}m_{3}}({\hat{k}}'')\nonumber \\  
 & \quad\quad  d^{i j}(\hat{z})e^{R}_{i j}(\hat{k})\, d^{k l}(\hat{z})e^{L}_{k l}(\hat{k})\, d^{m n}(\hat{z})e^{R}_{m n}({\hat{k}}')d^{k l}(\hat{z})e^{L}_{k l}({\hat{k}}') \phantom{\dfrac{1}{1}} \,. 
\label{tpangl}
\end{align}
For an isotropic, unpolarized SGWB, the coherency matrix is diagonal,
with $P_{RR}=P_{LL}$, so only the auto-power spectra $P_{RR}$ and
$P_{LL}$ contribute. For the explicitly written contribution in
\eqref{finaltpc}, this selects the helicity combinations $(L,R,R)$,
$(R,L,R)$, $(R,L,L)$, and $(L,R,L)$, while the remaining terms are
obtained by the corresponding permutations of the dummy variables. In deriving
the above equation, we have also used the identities
$e^{R*}_{ij}=e^L_{ij}$ and $e^{L*}_{ij}=e^R_{ij}$.

To evaluate the angular integrals in \eqref{tpangl}, we compute the contraction of the detector tensor with the circular polarization basis tensors. In our setup (coordinate system), where the pulsar is located along $\hat{z}$, it can be shown that
\begin{align}
d^{ij}(\hat{z})e^{R,L}_{i j}(\hat{k}) = & -\frac{\sin^{2}\theta}{\sqrt{2}} \nonumber \\
= & \, \frac{2}{3}\sqrt{\frac{2 \pi}{5}} Y_{2 0}(\hat{k})-\frac{2 \sqrt{2 \pi}}{3} Y_{0 0}(\hat{k}) \,.
\label{dpcon}
\end{align}
The second line expands the contraction in terms of spherical harmonics, which is convenient for subsequent angular integrations.
Following this, we now perform the angular integrations and summations over the left- and right-handed polarization states $A=L,R$ in the three-point autocorrelation function. The integral over ${\hat{k}}''$ yields
\begin{align}
\int\mathrm{d} {\hat{k}}'' Y_{l_{3}m_{3}}({\hat{k}}'')=\sqrt{4\pi}\delta_{l_{3}0}\delta_{m_{3}0}~.
\label{1stint}
\end{align}
To evaluate the integral over $\hat{k}$ in \eqref{tpangl}, we rewrite the integrand as a sum of products of three spherical harmonics:  
\begin{align}
\int \mathrm{d}\hat{k} \, Y_{l_{1}m_{1}}(\hat{k}) d^{i j}(\hat{z})e^{R}_{i j}(\hat{k}) d^{k l}(\hat{z})e^{L}_{k l}(\hat{k}) = \, & \frac{8 \pi}{45} \int \mathrm{d}\hat{k} ~Y_{l_{1}m_{1}}(\hat{k}) Y_{2 0}(\hat{k}) Y_{2 0}(\hat{k}) \nonumber \\ 
& -\frac{16 \pi}{9\sqrt{5}} \int \mathrm{d}\hat{k} ~Y_{l_{1}m_{1}}(\hat{k}) Y_{2 0}(\hat{k}) Y_{0 0}(\hat{k}) + \frac{8 \pi}{9} \int \mathrm{d}\hat{k} ~Y_{l_{1}m_{1}}(\hat{k}) Y_{0 0}(\hat{k}) Y_{0 0}(\hat{k})~. 
\label{kint}
\end{align}
The required integral identities for three scalar spherical harmonics are
\begin{align}
\int\mathrm{d}{\hat{k}}Y_{0 0}(\hat{k}) Y_{2 0}(\hat{k}) Y_{l_{1} m_{1}}(\hat{k})=\frac{\delta_{l_{1} 2}\delta_{m_{1} 0}}{\sqrt{4\pi} }~,
\label{inti1}
\end{align}
\begin{align}
\int\mathrm{d}{\hat{k}}Y_{0 0}(\hat{k}) Y_{0 0}(\hat{k}) Y_{l_{1} m_{1}}(\hat{k})=\frac{\delta_{l_{1}0}\delta_{m_{1} 0}}{\sqrt{4\pi} } \,,
\label{inti2}
\end{align}
and
\begin{align}
\int\mathrm{d}{\hat{k}}Y_{2 0}(\hat{k}) Y_{2 0}(\hat{k}) Y_{l_{1} m_{1}}(\hat{k})&=5\sqrt{\frac{(2l_{1}+1)}{4\pi}}\begin{pmatrix}
2 & 2 & l_{1}\\
0 & 0 & m_{1}
\end{pmatrix}
 \begin{pmatrix}
2 & 2 & l_{1}\\
0 & 0 & 0
\end{pmatrix} \nonumber \\ &= \delta_{m_{1} 0} \sqrt{\frac{(2l_{1}+1)}{4\pi}}~ \frac{5(l_{1}-3)^{2}(l_{1}-1)^{2}(l_{1}+2)^{2}(4+l_{1})^{2}}{ \Gamma(5-l_{1}) \Gamma(l_{1}+6)}
\label{inti3}
\end{align}
where \eqref{inti3} is nonzero only when the triangle conditions of the Wigner $3j$ symbols are satisfied; this means $l_1 \in \mathbb{Z}$ and $ 0 \le l_1 \le 4$.
In performing the sum over $l_{1}$, it is noted that the $2+2+l_{1}$ must be an even integer as in the Wigner 3j symbol, all $m$ are individually zero. Thus, the above condition leaves us with three choices of $l_{1}$ corresponding to $l_{1}=0, 2, 4$. For this condition of $l_{1}$, we are able to write down \eqref{inti3} for different value of $l_{1}$ as
\begin{align}
\int\mathrm{d}{\hat{k}}Y_{2 0}(\hat{k}) Y_{2 0}(\hat{k}) Y_{l_{1} m_{1}}(\hat{k})&= 
    \begin{cases}
    \delta_{m_{1} 0} / ( 2 \sqrt{\pi} )  \,, \quad & l_{1}=0 \\
    \delta_{m_{1} 0} \sqrt{5}/( 7 \sqrt{\pi} ) \,, \quad & l_{1}=2 \\ 
     \delta_{m_{1} 0} 3 / ( 7 \sqrt{\pi} ) \,, \quad & l_{1}=4 ~.   
     \end{cases}
\label{int3l}
\end{align}
Using the above identities, we evaluate \eqref{kint} as 
\begin{align}
\int \mathrm{d}\hat{k} \, Y_{l_{1}m_{1}}(\hat{k})& d^{i j}(\hat{z})e^{R}_{i j}(\hat{k}) d^{k l}(\hat{z})e^{L}_{k l}(\hat{k})=\delta_{m_{1}0}\left[\frac{8\sqrt{\pi}}{15}\delta_{l_{1}0}-\frac{16}{21}\sqrt{\frac{\pi}{5}}\delta_{l_{1}2}+\frac{8\sqrt{\pi}}{105}\delta_{l_{1}4}\right]~.
\label{compin}
\end{align}
For the remaining ${\hat{k}}'$ integration in \eqref{tpangl}, a similar expression to \eqref{compin} can be derived. The angular integrals over GW directions, $\hat{k}$, ${\hat{k}}'$ and ${\hat{k}}''$, imply that $m_{1}= m_{2}= m_{3}=0$ and $l_{3}=0$. Then, we can use the identity
\begin{align}
\begin{pmatrix}
l_{1} & l_{2} & 0\\
0 & 0 & 0
\end{pmatrix}=\delta_{l_{1}l_{2}} \frac{(-1)^{l_{1}}}{\sqrt{2l_{1}+1}}~.
\label{idenll}
\end{align}
By gathering all the previous integral identities, particularly \eqref{1stint}, \eqref{compin}, and \eqref{idenll}, we obtain
\begin{align}
\Upsilon (f,{f}',{f}'')=\frac{2}{\pi} \left[\frac{64 \pi}{225}J_{000}(f,{f}',{f}'')+\frac{\pi}{5}\left(\frac{16}{21}\right)^{2}J_{220}(f,{f}',{f}'')  +\pi\left(\frac{8}{105}\right)^{2}J_{440}(f,{f}',{f}'')\right] \,,
\label{finangc}
\end{align}
which is a function of GW frequencies. Useful analytic and numerical properties of $\Upsilon(f,f',f'')$ are discussed in Appendix \ref{sec:3pacf_Upsilon}.
Furthermore, it can be shown that $J_{ll0}(f, f', f'')$ can be computed analytically and takes the form
\begin{align}
	J_{ll0}(f, f', f'') = \frac{\pi \beta(\Delta (f, f', f''))}{4 f f' f''} P_{l}( \Delta (f, f', f'') )\,.
\label{bessel}
\end{align}
where $\beta(\Delta)$ is given by
\begin{align}
\label{eq:beta_def}
	\beta(\Delta) =
    \begin{cases}
        1 \,, & \text{for } -1<\Delta<1 \\
        1/2 \,, & \text{for } \Delta=\pm 1 \\
        0 \,, & \text{otherwise} \,,
    \end{cases}
\end{align}
and $\Delta$ is defined as
\begin{align}
    \Delta (f, f', f'') = \frac{f^{2}+f'^{2}-f''^{2}}{2 f f'} \,.
\end{align}
Note that $\Upsilon (-f,-{f}',-{f}'')=\Upsilon (f,{f}',{f}'')$, following symmetry property of the spherical Bessel functions under a change of sign of the frequency, $j_{l}(-f x)=j_{l}(fx)$ for even values of $l$.

The complete positive-frequency reduction of ~\eqref{tpfcr} includes all (eight) sign sectors. Since the Bessel orders entering $\Upsilon$ are even, its Bessel-integral definition is even in each frequency separately. Using $c(-f,t,\eta_r)=c^{*}(f,t,\eta_r)$, the full expression is
\begin{align}
\left \langle r(t,\hat{e})^{3} \right\rangle
\sim {}& 2g_{\rm NL}\left(\frac{i}{4\pi}\right)^3
\int_0^\infty f\,\mathrm{d}f
\int_0^\infty f'\,\mathrm{d}f'
\int_0^\infty f''\,\mathrm{d}f''\,
\bigl[P_{LL}(f)P_{RR}(f')+P_{RR}(f)P_{LL}(f')\bigr]
\Upsilon(f,f',f'') \nonumber\\
&\quad\times[c(f)-c^{*}(f)]\,[c(f')-c^{*}(f')]\,[c(f'')-c^{*}(f'')]
+\{f'\leftrightarrow f''\}+\{f\leftrightarrow f''\} \,,
\label{eq:mu3_full_sign_sectors}
\end{align}
where the common arguments $(t,\eta_r)$ of $c$ have been suppressed.
With $\eta_r=0$, the full phase product reduces to
\begin{align}
[c(f)-c^{*}(f)]\,[c(f')-c^{*}(f')]\,[c(f'')-c^{*}(f'')]
=-8i\sin(2\pi ft)\sin(2\pi f't)\sin(2\pi f''t).
\label{eq:mu3_full_sign_sines}
\end{align}

Within this detector-projected phenomenological benchmark, the timing-residual third moment is nonzero and polarization-frame independent.
This provides an explicit polarization-frame-independent
detector-projected construction with a nonvanishing timing residual
skewness. It should be regarded as an alternative phenomenological
construction to the helicity-space ansatz considered in
Section \ref{subsec:no-go}.

We emphasize that the above result should be considered as approximate, or an order-of-magnitude estimate, subject to the approximations that were enumerated earlier. We acknowledge this by only using the asymptotic symbol `$\sim$' in every time we expressed $\langle r(t, \hat{e})^3 \rangle$ in this section.

\section{Numerical estimates}
\label{sec:testing_gaussianity}

We compare the third moment of the detector-projected benchmark with the Gaussian ensemble rms of a single-epoch cubic residual. Section~\ref{subsec:noise} derives this normalization; Section~\ref{subsec:signal} presents the existing numerical benchmark.

\subsection{
Gaussian ensemble variance of the cubic residual
}
\label{subsec:noise}
We compare Gaussian and non-Gaussian timing residual distributions through their moments. Throughout this comparison, $r$ has zero ensemble mean. Its dimensional third moment is $\mu_3(t,\hat e)\equiv\langle r(t,\hat e)^3\rangle$, with units of time cubed; the dimensionless skewness is defined below.

For a zero-mean Gaussian residual, $\mu_3=0$, but the random variable ${\cal S}=r(t,\hat e)^3$ has a nonzero ensemble variance. This is the scatter of a single-epoch cubic residual, not the variance of a sample-averaged skewness estimator.
For one pulsar at a fixed epoch, the ensemble variance of ${\cal S}$ is
\begin{align}
\sigma^{2} & \equiv \langle {\cal S}^2 \rangle - \langle {\cal S} \rangle^2 \nonumber \\
& =\langle r(t,\hat{e})^{6}\rangle-\langle r(t,\hat{e})^{3}\rangle^2 \,.
\label{Eq:sigma2}
\end{align}
The third moment of the zero-mean Gaussian residual vanishes,
\begin{align}
\langle r(t,\hat{e})^{3}\rangle=0\,.
\label{skewness}
\end{align}
Applying Isserlis' theorem \cite{1918Biome..12..134I}, the six-point autocorrelation function for the same pulsar becomes
\begin{align}
\langle r(t,\hat{e})^{6}\rangle=15 \langle r(t,\hat{e})^{2}\rangle^3~.
\label{Eq:r6}
\end{align}

Following the approximations that were done to calculate \eqref{eq:mu3_full_sign_sectors}, we arrive at a familiar expression for the corresponding timing residual two-point autocorrelation function:\footnote{An identity that we found useful in deriving this expression starting with the detector-projected GW statistics is
\begin{equation}
\int_{0}^{\infty} \mathrm{d}x \, x^2  \, j_0(f x) \, j_0({f}'x) =\frac{\pi}{2f{f}'}\left(\delta(f-{f}')-\delta(f+{f}')\right)\,.
\label{bessel_iden}
\end{equation}
}
\begin{equation}
\label{eq:timing_residual_variance}
    \langle r(t,\hat{e})^2 \rangle \sim \int_{0}^{\infty}\mathrm{d}f \, c(f,t,\eta_r)  \, c^{*}(f,t,\eta_r) \, S(f) \,,
\end{equation}
where the one-sided power spectral density (PSD) $S(f)$ is identified with the SGWB power spectra $P_{RR}(f)$ and $P_{LL}(f)$ as
\begin{equation}
\label{eq:P_to_S}
    S(f) = \dfrac{1}{15 \pi^{3/2}} \left( P_{RR}(f) + P_{LL}(f) \right) \,.
\end{equation}
When generalized to $t \neq t'$, the result can be taken as a statement of the Wiener-Khinchin theorem for a stationary, Gaussian signal at leading-order.
The Gaussian ensemble variance of the cubic residual is therefore
\begin{align}
\label{eq:r3_gaussian_noise}
    \sigma(t, \hat{e})^2 \sim 15 \left( \int_{0}^{\infty}\mathrm{d}f \, c(f,t,\eta_r)  \, c^{*}(f,t,\eta_r) \, S(f) \right)^3\,.
\end{align}
We approximate the observational frequency selection by an amplitude transmission function ${\cal T}(f)$ equal to one for $f_{\min}<|f|<f_{\max}$ and zero elsewhere. The covariance contains $|{\cal T}(f)|^2$, which equals the same top-hat. The Gaussian cubic variance becomes
\begin{align}
\label{eq:r3_gaussian_noise_tophat}
    \sigma(t, \hat{e})^2 \sim 15 \left( \int_{f_{\rm min}}^{f_{\rm max}}\mathrm{d}f \, c(f,t,\eta_r)  \, c^{*}(f,t,\eta_r) \, S(f) \right)^3\,.
\end{align}

\subsection{
Gaussian-rms-normalized timing residual skewness
}
\label{subsec:signal}

\begin{figure}[t]
    \centering
    \includegraphics[width=0.6\linewidth]{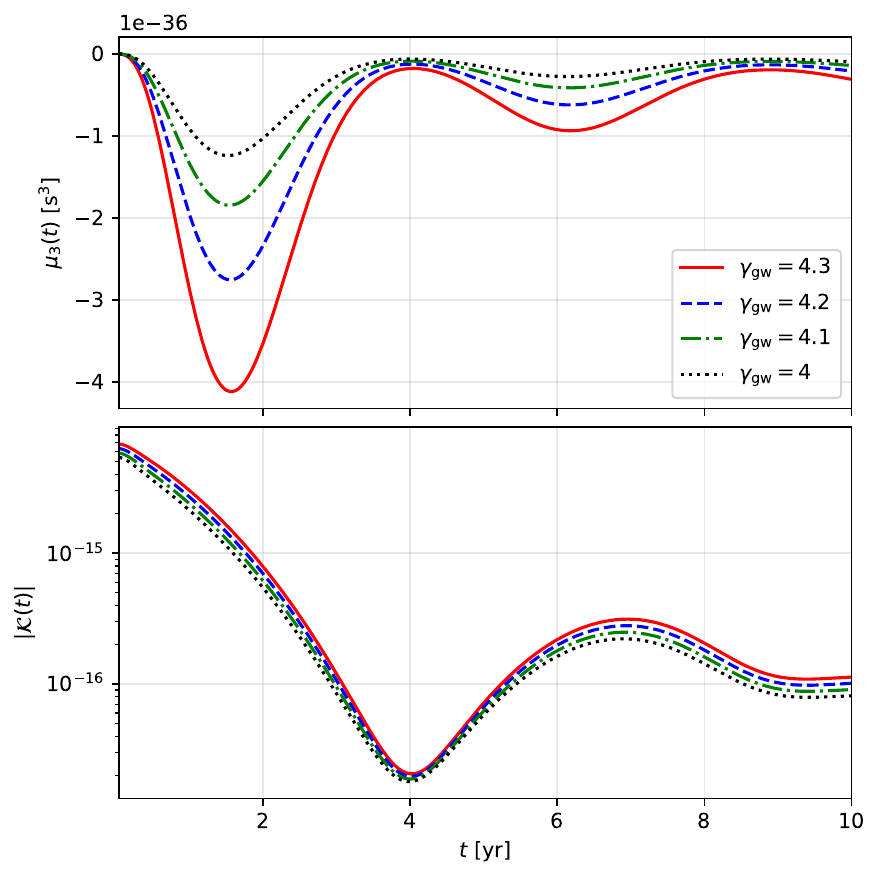}
    \caption{
        Existing detector-projected benchmark: the upper panel shows the third moment $\mu_3(t)$ in ${\rm s}^3$; the lower panel shows $|\mathcal K(t)|$, the Gaussian-rms-normalized skewness \eqref{eq:normalized_skewness}.
        Curves use $g_{\rm NL}=1$, $A_{\rm gw}=10^{-15}$, $\gamma_{\rm gw}=4.3,4.2,4.1,4.0$, and $P_{RR}=P_{LL}$. The legend's power laws indicate spectral slopes, with normalization given by \eqref{eq:psd_power_law}. The epoch $t$ varies within a nominal ten-year baseline.
        }
    \label{fig:r3_alt_varspec_results}
\end{figure}

We now present the existing evaluation of \eqref{eq:mu3_full_sign_sectors} as a phenomenological benchmark.
The existing numerical calculation restricts the two power spectrum legs to the nominal PTA band, but extends the remaining frequency to the triangle-supported range.
The integrals over $f$ and $f'$ which are weighted by the GW power spectra $P_{RR}(f)$ and $P_{LL}(f)$ have been truncated to the PTA sensitivity range.  In the permutation terms, the frequency labels are
interchanged together with their corresponding integration ranges.
Thus, in each term, the two frequencies entering the GW power spectra
are integrated over $[f_{\min},f_{\max}]$, while the remaining
frequency is integrated over $[0,2f_{\max}]$ and is further restricted
by the support of $\Upsilon$.
In particular, for the explicitly written
term in \eqref{eq:mu3_full_sign_sectors}, the integration range of $f''$ is constrained by the
condition $-1<\Delta<1$ in \eqref{eq:beta_def}, or a triangle inequality condition on the three frequencies $f$, $f'$, and $f''$:
\[
|f-f'| < f'' < f+f'.
\]
This restriction originates in the momentum-conserving Dirac delta function appearing in the three-point correlation function of the detector-projected GW amplitude.

Powell and Tasinato \cite{Powell:2019kid} study stationary non-Gaussianity with folded momentum configurations. Tsuneto et al.\ \cite{Tsuneto:2018tif} construct filters that select momentum-triangle configurations. Our equal-time channel benchmark integrates over triangle-supported frequencies without imposing the stationary signed-frequency constraint.

We distinguish the third moment, dimensionless skewness, and Gaussian-rms-normalized skewness:
\begin{align}
\mu_3(t,\hat e)&\equiv\langle r(t,\hat e)^3\rangle,\nonumber\\
\gamma_1(t,\hat e)&\equiv\frac{\mu_3(t,\hat e)}{\langle r(t,\hat e)^2\rangle^{3/2}},\nonumber\\
\mathcal K(t,\hat e)&\equiv\frac{\mu_3(t,\hat e)}{\sqrt{15\langle r(t,\hat e)^2\rangle^3}}
=\frac{\gamma_1(t,\hat e)}{\sqrt{15}}.
\label{eq:normalized_skewness}
\end{align}
The covariance in the normalization is evaluated at Gaussian order, consistently with the leading-order numerator. The quantities $\gamma_1$ and $\mathcal K$ are dimensionless and are defined where the variance is nonzero. In particular, the normalization is undefined at the reference epoch $t=0$, where $r=0$.
The denominator of $\mathcal K$ is $\sigma$ from \eqref{eq:r3_gaussian_noise_tophat}, the Gaussian rms of ${\cal S}=r^3$. It is not an uncertainty obtained by averaging epochs or combining pulsars. For a Gaussian residual $\mathcal K=0$. For the benchmark we assume equal helicity power, $P_{RR}(f)=P_{LL}(f)$, and a smooth power-law effective one-sided PSD. We set $g_{\rm NL}=1$ as a normalization, not a universal upper bound; other couplings rescale the leading-order result while the perturbative approximation remains valid. The PSD is
\begin{equation}
\label{eq:psd_power_law}
    S(f) \equiv \frac{A_{\rm gw}^2}{12 \pi^2} \left( \dfrac{f}{f_{\rm yr}} \right)^{-\gamma_{\rm gw}} f_{\rm yr}^{-3} \,,
\end{equation}
which determines the timing residual variance through \eqref{eq:timing_residual_variance}; \eqref{eq:P_to_S} gives the associated helicity power spectra. At $f_{\rm yr}=1\,{\rm yr}^{-1}\simeq31.8\,{\rm nHz}$, the PSD is $A_{\rm gw}^2f_{\rm yr}^{-3}/(12\pi^2)\simeq2.6\times10^{-10}(A_{\rm gw}/10^{-15})^2\,{\rm s}^3$. A circular, GW-driven SMBHB population gives the familiar reference slope $\gamma_{\rm gw}=13/3$ \cite{Phinney:2001di, Bernardo:2024bdc}; choosing a nearby slope does not assign that origin to our model. The retained curves use $A_{\rm gw}=10^{-15}$, slopes $4.3,4.2,4.1,4.0$, a nominal baseline $T=10\,{\rm yr}$ and power spectrum band $[1/T,14/T]$. The horizontal coordinate $t$ is the elapsed epoch within this fixed baseline. The numerical benchmark fixes $\eta_r=0$. A realistic timing-model projection is outside this calculation \cite{Hazboun:2019vhv, Pitrou:2024scp, Allen:2025waa}.

At fixed spectral shape, response, band, and epochs, \eqref{eq:mu3_full_sign_sectors}, \eqref{eq:P_to_S}, and \eqref{eq:psd_power_law} imply
\begin{equation}
\mu_3\propto g_{\rm NL}A_{\rm gw}^{4},\qquad
\langle r^2\rangle^{3/2}\propto A_{\rm gw}^{3},\qquad
\gamma_1,\mathcal K\propto g_{\rm NL}A_{\rm gw}.
\label{eq:skewness_amplitude_scaling}
\end{equation}
Thus the small strain amplitude supplies a natural suppression of normalized skewness for the unit-coupling benchmark. The coefficient and time dependence depend on the chosen model and frequency treatment.

The retained benchmark in Figure~\ref{fig:r3_alt_varspec_results} has a third moment of order $10^{-36}\,{\rm s}^3$, with sign changes as the epoch varies. Within the illustrated range, redder spectra increase its magnitude. These values describe the projected-channel model and its existing frequency prescription.

Away from the zeros of the third moment, the retained curves reach $|\mathcal K|$ of order $10^{-15}$. This compares an ensemble third moment with the rms of a single-epoch cubic Gaussian residual. We have not evaluated the covariance of a non-Gaussian estimator or combined measurements across epochs and pulsars.

The benchmark also retains only the Earth term and replaces the exact PTA antenna response by a simplified detector projection. Familiar two-point results do not establish the correction to a three-point auto-statistic, where response products and cancellations differ. Neither correction has been calculated, and neither is assumed to be bounded by an order-one factor.
Restoring the exact antenna response and pulsar terms would provide a more complete calculation of the channel statistic.
That calculation, together with an estimator covariance, is left for future work.

%%%%%%%%%%%%%%%%%%%%%%%%%%%%%
\section{Conclusions}\label{conclu}
%%%%%%%%%%%%%%%%%%%%%%%%%%%%%%%

Motivated by CMB searches for weak primordial non-Gaussianity \cite{Komatsu:2001rj, Komatsu:2003iq}, we have investigated a perturbative ansatz for GW statistics and its timing residual three-point function. The primordial motivation concerns the information carried by higher moments beyond the leading covariance. Our analysis has yielded two main results. First, the
perturbative helicity-space ansatz (\ref{eq:hNL_sum}-\ref{eq:hnl_general}), as written,
does not produce a manifestly polarization-frame-invariant
timing residual bispectrum for tensor and vector GW polarizations. Second, a nontrivial construction of a non-Gaussian signal formulated in terms of the
detector-projected GW amplitudes makes the cancellation of the
polarization-frame phases explicit and yields a nonzero
timing residual skewness.

First, we showed that the tensor and vector response coefficients
carry spin-dependent phases under a change of polarization basis
(Section~\ref{subsec:no-go}). For the PTA timing residual bispectrum
to be polarization-frame independent, the helicity bispectrum in
\eqref{eq:ensemble_average_g_pert} must carry the corresponding
compensating phases. Achieving this may require either a suitable
generalization of the non-Gaussian ansatz for the GW Fourier
amplitudes or, while retaining its basic structure, the introduction
of the required spin-dependent factors directly into the resulting
helicity bispectrum. Determining these factors and incorporating them
into a polarization-frame-covariant formulation is left for future
work.

To address this limitation, we proposed a construction in which
non-Gaussian statistics are introduced in the detector-projected GW
amplitude (Section~\ref{subsec:three_point_correlation_nonGaussian}),
rather than in the GW amplitude itself. This approach is motivated by
the operational principle of GW detection, according to which the
measured signal is determined by the projection of the GW field onto
the detector response. In this construction, the helicity amplitude
and the polarization tensor carry opposite spin phases, so that their
product is polarization-frame invariant. Consequently, the resulting
three-point correlation function of the timing residuals is also
polarization-frame independent. By introducing non-Gaussianity at the
level of the detector response, we derived a nontrivial three-point
autocorrelation function, (skewness) of the timing residuals.

For the existing unit-coupling benchmark with $A_{\rm gw}=10^{-15}$, the third moment is of order $10^{-36}\,{\rm s}^3$ and its Gaussian-rms-normalized magnitude reaches order $10^{-15}$. The scaling $\mathcal K\propto g_{\rm NL}A_{\rm gw}$ explains why a small normalized third moment is natural in this perturbative model.

The primordial motivation remains relevant. A consistent higher-point calculation can test information absent from a leading-order power spectrum. However, full-sky integration and statistical isotropy do not identify a primordial source, and non-Gaussianity alone does not imply nonzero skewness. A measured third moment would probe phase-coupled statistics whose source interpretation requires a physical model. Our benchmark therefore does not establish that a detectable skewness must be astrophysical or that primordial non-Gaussianity is inaccessible to PTAs.

The next theoretical step is a detector-independent, spin-covariant primordial tensor construction \cite{Bartolo:2018qqn}. Quantitative PTA forecasts additionally require the exact response, pulsar terms, a timing-model projection, and the covariance of a specified estimator. Higher connected moments, including the four-point function \cite{Ciprini:2026pvz, Kuntz:2026usl}, can probe non-Gaussianity even when the bispectrum vanishes.

\section{Acknowledgments}
This work was supported by the National Science and Technology Council (NSTC) of Taiwan, Republic of China, under Grant Nos. NSTC 114-2811-M-008-024 (R.M.) and NSTC 115-2112-M-001-004 (K.W.N.).

\appendix

\section{Two point function: the Hellings \& Downs curve and beyond}
\label{sec:twopoint_hd}

In this section, we derive the timing residual correlation between a pair of Galactic pulsars. This is the leading PTA GW signal due to a nanohertz SGWB.

\subsection{Real space formalism}
\label{subsec:two_point_rsf}

The relevant observable is the timing residual correlation:
\begin{equation}
\label{eq:timing_residual_correlation_def}
    \langle r\left(t, \hat{e}_a\right) r\left(t', \hat{e}_b\right) \rangle \,.
\end{equation}
This describes the correlation between timing measurements between a pair of pulsars.

Using \eqref{eq:timing_residual_gw_signal_rsf}, we obtain
\begin{equation}
\label{eq:twopoint_rsf_general}
\begin{split}
    \langle r\left(t, \hat{e}_a\right) r\left(t' \hat{e}_b\right) \rangle = \sum_{AA'} & \int_{-\infty}^\infty \dd f \, C(f, t, t') \\
    & \times \int_{\mathbb{S}^2} \dd\hat{k} \dfrac{P_{AA'}\left(f, \hat{k}\right)}{(2\pi)^2} F^A\left(\hat{k}, \hat{e}_a, v\right) U\left(f D_a, \hat{k} \cdot \hat{e}_a, v\right) F^{A'*}\left(\hat{k}, \hat{e}_b, v\right) U^*\left(f D_b, \hat{k} \cdot \hat{e}_b, v\right) \\
    & \,\,\,\, + {\cal O}\left(\gNL^2\right) \,,
\end{split}
\end{equation}
where $C(f, t, t')$ is a time kernel given by
\begin{equation}
\label{eq:time_kernel}
    C(f, t,t')=(1-e^{-2\pi i f t})(1-e^{2\pi i f t'}) \,,
\end{equation}
and we have used the assumption that each GW polarization and mode satisfies the statistics of (\ref{eq:hNL_sum}-\ref{eq:hnl_general}); i.e.,
\begin{equation}
\label{eq:gw_2pf_fourier_space}
    \blangle h_A\left(f, \hat{k}\right) h_{A'}^* \left( f', \hat{k}'\right) \brangle = \frac{1}{f^2}\delta \left( f - f' \right) \delta^{2} \left( \hat{k} - \hat{k}' \right) P_{AA'} \left(f, \hat{k}\right) + {\cal O}\left( \gNL^2 \right) \,.
\end{equation}

In the isotropic case, and when the GW polarizations do not interfere with one another, $P_{AA'}(f, \hat{k}) = Y_{00}(\hat{k})\delta_{AA'} P_{AA}(f)$, the result simplifies to
\begin{equation}
\label{eq:twopoint_rsf_isotropic}
\begin{split}
    \langle r\left(t, \hat{e}_a\right) r\left(t' \hat{e}_b\right) \rangle = \sum_{A} & \int_{-\infty}^\infty \dd f \, C(f, t, t') \dfrac{P_{AA}\left(f\right)}{4 \pi^2 } \gamma^A_{ab}(f) + {\cal O}\left(\gNL^2\right) \,,
\end{split}
\end{equation}
where the overlap reduction function, $\gamma_{ab}^A\left( f \right)$, is defined as
\begin{equation}
\label{eq:orf_realspace}
    \gamma_{ab}^A\left( f, v \right) = \int_{\mathbb{S}^2} \dfrac{ \dd\hat{k} }{\sqrt{4\pi}} F^A\left(\hat{k}, \hat{e}_a, v\right) U\left(f D_a, \hat{k} \cdot \hat{e}_a, v\right) F^{A*}\left(\hat{k}, \hat{e}_b, v\right) U^*\left(f D_b, \hat{k} \cdot \hat{e}_b, v\right) \,.
\end{equation}
In GR, the ORF reduces to the HD curve. The result \eqref{eq:twopoint_rsf_isotropic} could be viewed as a statement of stationarity of the signal (or the Wiener-Khinchin theorem), until ${\cal O}\left(\gNL^2\right)$ corrections become detectable.

\subsection{Harmonic space formalism}
\label{subsec:two_point_harmonic_space}

We compute the two-point function \eqref{eq:timing_residual_correlation_def} in harmonic space, utilizing \eqref{eq:timing_residual_harmonic_coefficients}. In this case, we obtain
\begin{equation}
    \langle r\left(t, \hat{e}_a\right) r\left(t', \hat{e}_b\right) \rangle = \sum_{l m} \sum_{l' m'} \langle a_{lm}(t) a^*_{l'm'}(t') \rangle Y_{lm}\left( \hat{e}_a \right) Y^*_{l'm'}\left( \hat{e}_b \right) + {\cal O}\left( \gNL^2 \right) \,,
\end{equation}
where
\begin{equation}
\label{eq:two_point_lm}
    \langle a_{lm}(t) a^*_{l'm'}(t') \rangle = \int_{-\infty}^\infty \dd f \, C(f, t, t') \sum_{A A'} \int_{\mathbb{S}^2} \dd\hat{k} \ P_{AA'}\left( f, \hat{k} \right) J^{A}_{lm} \left( f D_a, v, \hat{k} \right) J^{A' *}_{l'm'} \left( f D_b, v, \hat{k} \right) \,,
\end{equation}
and
\begin{equation}
\label{eq:Jlm_def}
    J_{lm}^A \left( fD, v, \hat{k} \right) = \int_0^{2\pi f D v} \dfrac{\dd x}{v} \ e^{i x/v} \sum_{LM} i^L Y^*_{LM} \left( \hat{k} \right) j_L(x) \int_{\mathbb{S}^2} \dd\hat{e} \ \left[ d^{ij}(\hat{e})\, e_{ij}^A\left(\hat{k}\right) \right] Y_{LM}\left( \hat{e} \right) Y_{lm}^*\left(\hat{e}\right) \,.
\end{equation}
The $J_{lm}^A(fD, v, \hat{n})$'s are the generalization to the projection factors \cite{Qin:2018yhy, Qin:2020hfy} for arbitrary GW speed and finite pulsar distances.

In the isotropic scenario and without interference between GW polarizations, it can be shown that the timing residual correlation reduces to \eqref{eq:twopoint_rsf_isotropic} where the ORF is given by
\begin{equation}
\label{eq:orf_harmonic space}
    \gamma_{ab}^A \left( f, v \right) = 4\pi^2 \sum_{lm} \sum_{l'm'} Y_{lm}\left( \hat{e}_a \right) Y^*_{l'm'} \left( \hat{e}_b \right) \int_{\mathbb{S}^2} \dd\hat{k} \ Y_{00}\left(\hat{k}\right) J^A_{l m} \left( f D_a, v, \hat{k} \right) J^{A*}_{l'm'} \left( f D_b, v, \hat{k} \right) \,.
\end{equation}
Note that $Y_{00}\left(\hat{k}\right) = 1/\sqrt{4\pi}$.

The real space and harmonic space expressions of the ORF have been explicitly shown to be equivalent in GR and beyond \cite{Mingarelli:2013dsa, Qin:2018yhy, Qin:2020hfy, Liu:2022skj, Bernardo:2022rif}.

\section{Deriving the generalized projection factors}
\label{sec:generalized_projection_factors}
To evaluate this quantity, we choose the propagation direction of the gravitational wave $\hat{k}$ to lie along the polar (or $\hat{z}$) axis. Then we specify that the pulsar is located in the direction
\begin{align}
    \hat{e}
= \sin\theta \cos\phi\, \hat{x}
+ \sin\theta \sin\phi\, \hat{y}
+ \cos\theta\, \hat{z} \, .
\end{align}
This choice implies the simplification
\begin{align}
 Y_{L M}\left(\hat{k}=\hat{z}\right) \equiv \sqrt{\frac{2L+1}{4\pi}} \delta_{M0} \,.  
\end{align}
Then, the contraction between the detector tensor $d^{ij}(\hat{e})$ and the polarization tensor gives 
\begin{align}
    d^{ij}(\hat{e}) e^{R,\,L}_{ij}(\hat{z})=4\sqrt{\frac{\pi}{15}} Y_{2\pm 2}(\hat{e}) \,,
\end{align}
where the helicity index takes the value $+2$ for $R$ polarization and $-2$ for $L$ polarization. Using these relations and performing the angular integration over $\dd \hat{e}$ from \eqref{redshift_fluc}, we obtain the redshift fluctuation for a pulsar induced by a GW propagating along the $z$-direction as 
\begin{align}
J^{R,\, L}_{l\,m}\left(fD,v,\hat{z}\right)=-\delta_{m\pm 2}\, \sqrt{\frac{\left(2l+1\right)}{8\pi}}\, \left(i^l \sqrt{\frac{\left(l+2\right)!}{\left(l-2\right)!}} \int_{0}^{2\pi f D v} \dfrac{\dd x}{v} \, e^{ix/v} \,\dfrac{j_{l}(x)}{x^2} \right)\,.
\label{JAlmz}
\end{align}
To generalize the result, we  reorient the $\hat{z}$ into an arbitrary direction $\hat{k}=(\theta, \phi)$. This is done by acting a rotation operator on the last result:
\begin{align}
J^{A}_{l\,m}\left(fD,v,\hat{k}\right)= \sum_{m'}D^{l\,*}_{m'\,m}(-\alpha,\, -\theta, \,-\phi)J^{A}_{l\,m'}\left(fD,v,\hat{z}\right) \,,
\end{align}
where Wigner-$D$ matrix is given by
\begin{align}
D^{l}_{m'\,m}(-\alpha,-\theta,-\phi)
=
\sqrt{\frac{4\pi}{2l+1}}\;
{}_{-m'}Y_{l m}(\theta,\phi)\, e^{i m' \alpha}\,.
\end{align}
Above, ${}_{s}Y_{lm}(\hat{k})$'s are known as spin-weighted or spin-$s$ spherical harmonics; the standard $Y_{lm}(\hat{k})$ corresponds to the spinless case $s = 0$.
Then, we have
\begin{align}
J^{R}_{l\,m}\left(fD,v,\hat{k}\right)&= D^{l\,*}_{2\,m}(-\alpha,\, -\theta, \,-\phi)J^{R}_{l\,2}\left(fD,v,\hat{z}\right) \nonumber \\
J^{L}_{l\,m}\left(fD,v,\hat{k}\right)&=D^{l\,*}_{-2\,m}(-\alpha,\, -\theta, \,-\phi)J^{L}_{l\,-2}\left(fD,v,\hat{z}\right) \,.
\end{align}
The phase factor $e^{i m'\alpha}$ reflects the remaining freedom to
rotate the polarization basis about the propagation direction $\hat{k}$. Consequently, the response coefficients $J^A_{lm}$ acquire polarization-frame
dependent spin phases. These phases do not affect physical observables such as the two-point function
when they are accompanied by the corresponding opposite spin phases
of the GW polarization quantities. In the polarized
two-point function, the phases carried by the response coefficients
are canceled by those of the corresponding spin-weighted spherical
harmonics \cite{Liu:2022skj}. The explicit application of the rotation matrices give \eqref{eq:Jlm_tensor}.

\section{Analytic and numerical properties of the three-point autocorrelation function}
\label{sec:3pacf_Upsilon}

 \begin{figure}[ht]
     \centering
     \includegraphics[width=0.475\textwidth]{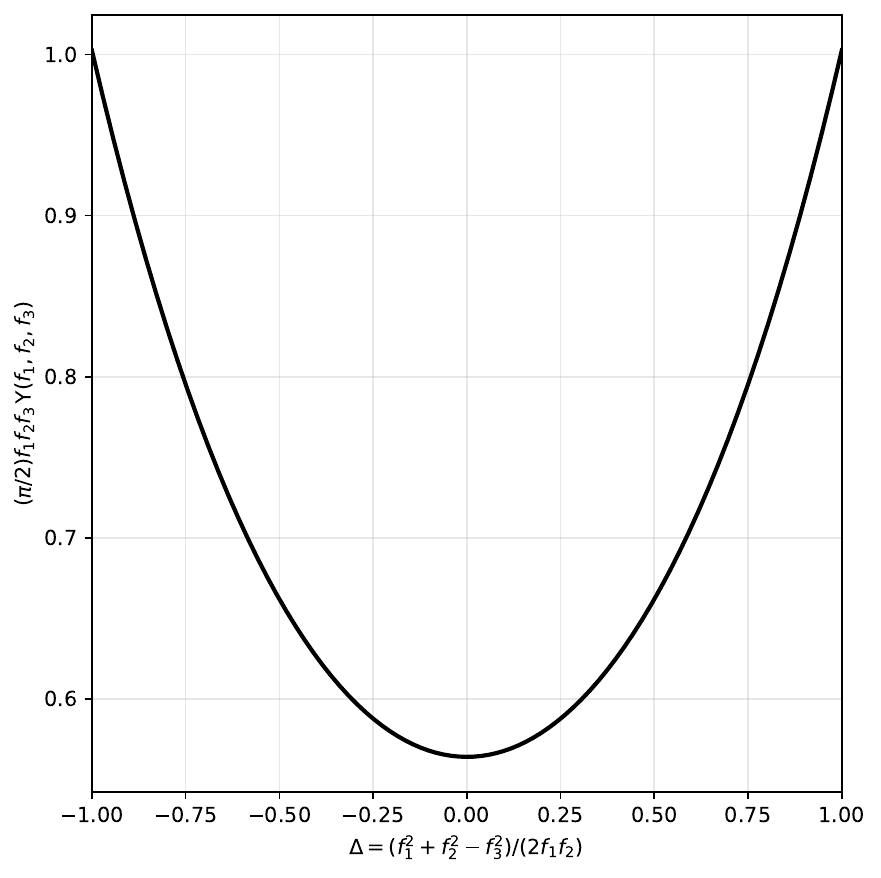}
     \caption{The normalized three-point autocorrelation function $X(\Delta) \equiv (\pi/2) f_1 f_2 f_3 \Upsilon(f_1, f_2, f_3)$ as a function of $\Delta \equiv \Delta(f_1, f_2, f_3) = (f_1^2 + f_2^2 - f_3^2)/(2 f_1 f_2)$. Outside the domain shown in this figure, the function value is zero. All values of, or the entire domain and range of, $\Upsilon(f_1, f_2, f_3)$ can be deduced based on this figure.}
     \label{fig:Upsilon_domain}
 \end{figure}

\begin{figure}[ht]
     \centering
     \includegraphics[width=0.475\textwidth]{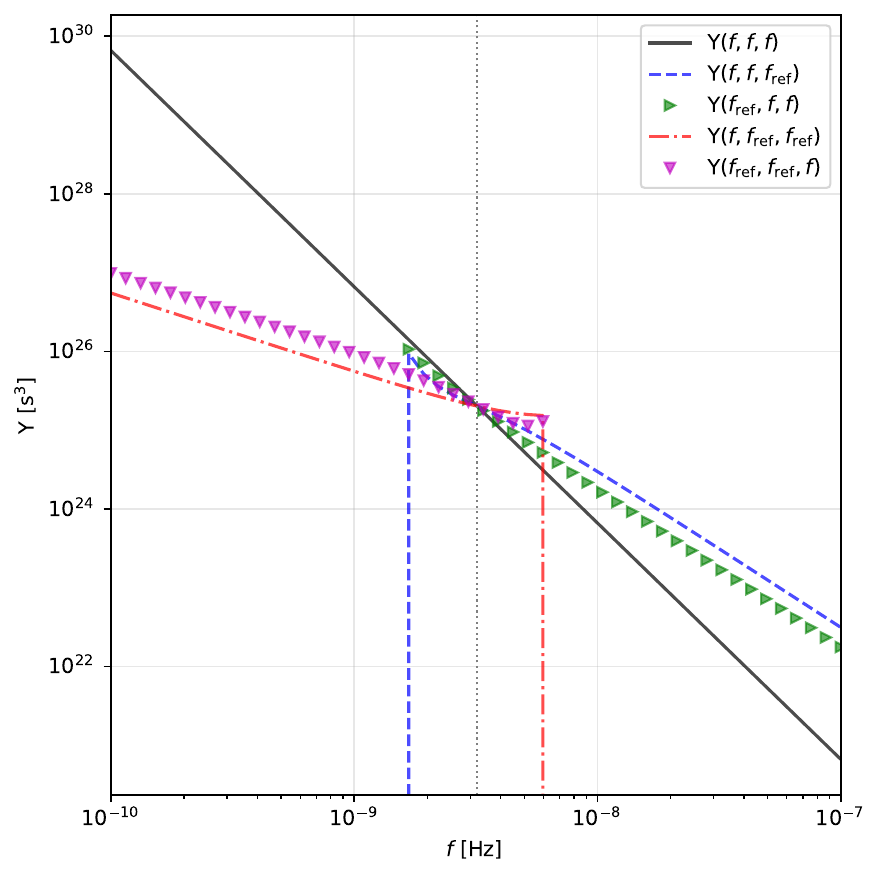}
     \caption{Diagonal and some off-diagonal components of $\Upsilon(f_1, f_2, f_3)$; with fixed parameters $f_{\rm min}=10^{-10} \, {\rm Hz}$, $f_{\rm max}=10^{-7} \, {\rm Hz}$, and $f_{\rm ref}=0.1 \, {\rm yr}^{-1} \sim 3.2 \times 10^{-9} \, {\rm Hz}$. The vertical dotted line shows the reference frequency $f_{\rm ref}$.}
     \label{fig:Upsilon_1d}
 \end{figure}

 \begin{figure}[ht]
     \centering
     \includegraphics[width=0.9\linewidth]{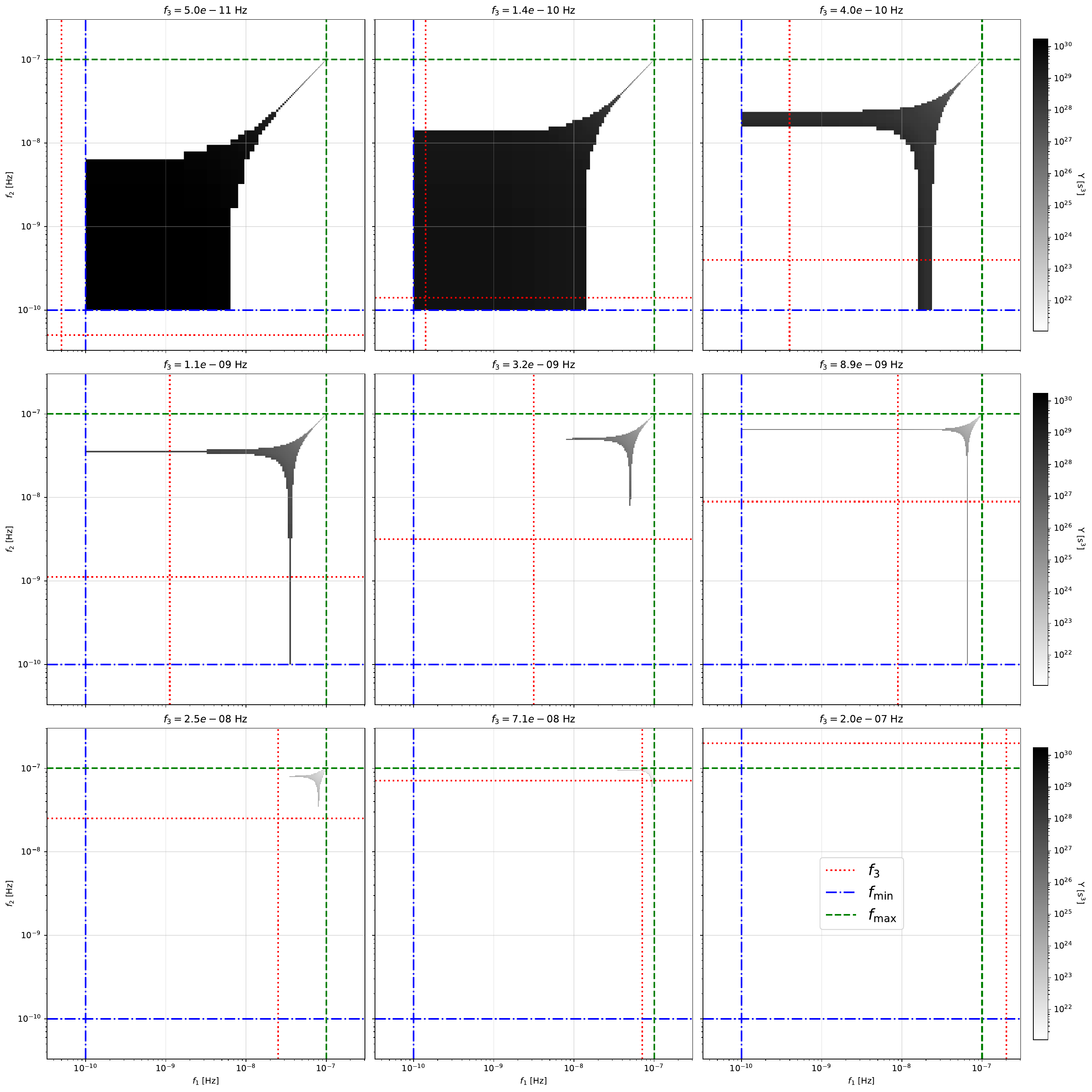}
     \caption{Off-diagonal components of $\Upsilon(f_1, f_2, f_3)$; with $f_{1,2} \in [f_{\rm min}, f_{\rm max}]$, fixed parameters $f_{\rm min}=10^{-10} \, {\rm Hz}$, $f_{\rm max}=10^{-7} \, {\rm Hz}$, and fixed $f_3 \in [f_{\rm min}, f_{\rm max}]$ values per subplot (shown at the top of each subplot). In each subplot, the horizontal axis is $f_1$ and the vertical axis $f_2$. All subplots are normalized relative to the same scale for visual comparison with each one.
     The red dotted line shows the fixed $f_3$ values relative to both axes, the green dashed line $f_{\rm max}$, and the blue dash-dotted line $f_{\rm min}$.}
     \label{fig:Upsilon_f1f2_vs_f3}
 \end{figure}

\begin{figure}[ht]
     \centering
     \includegraphics[width=0.9\linewidth]{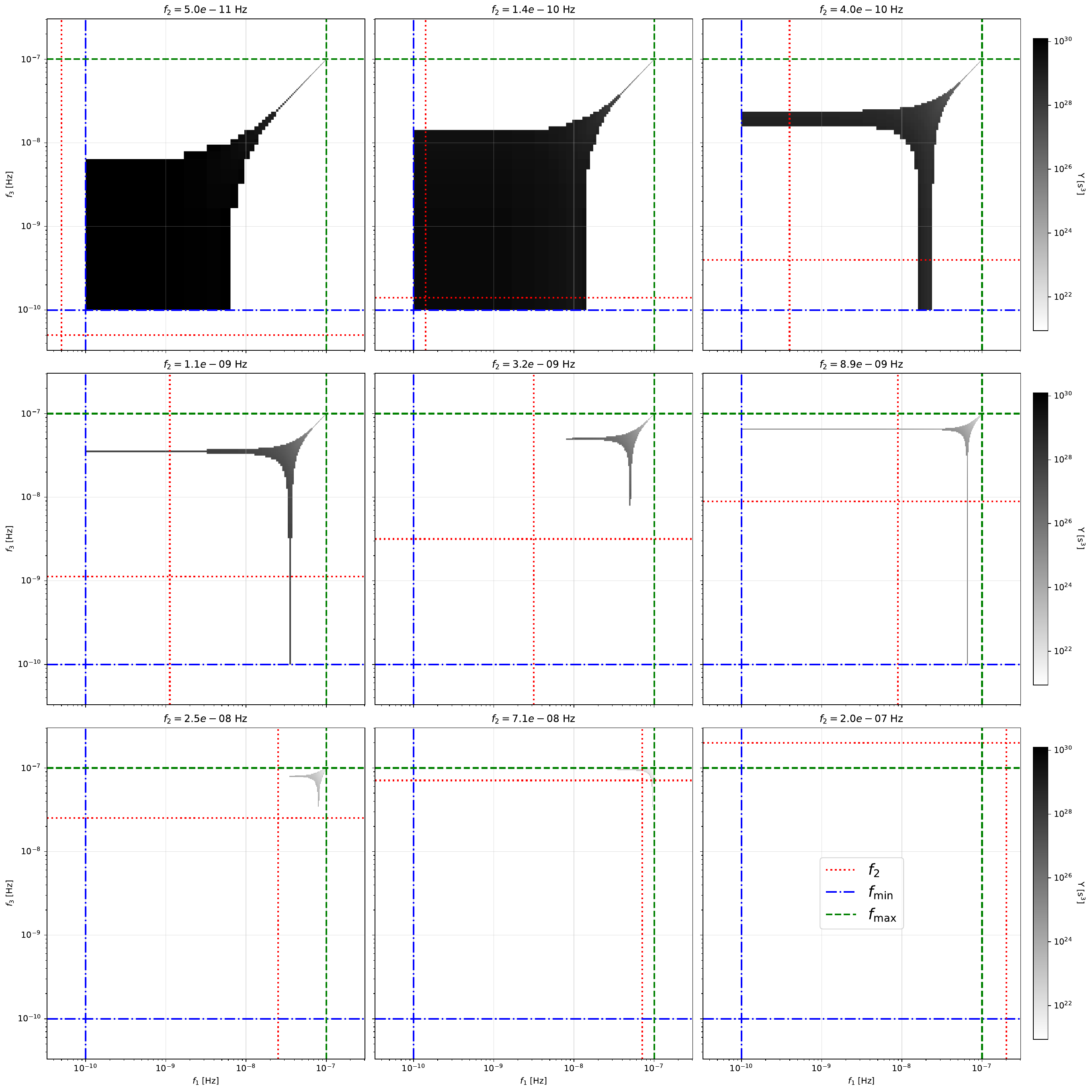}
     \caption{Off-diagonal components of $\Upsilon(f_1, f_2, f_3)$; with $f_{1,3} \in [f_{\rm min}, f_{\rm max}]$, fixed parameters $f_{\rm min}=10^{-10} \, {\rm Hz}$, $f_{\rm max}=10^{-7} \, {\rm Hz}$, and fixed $f_2 \in [f_{\rm min}, f_{\rm max}]$ values per subplot (shown at the top of each subplot). In each subplot, the horizontal axis is $f_1$ and the vertical axis $f_3$. All subplots are normalized relative to the same scale for visual comparison with each one.
     The red dotted line shows the fixed $f_2$ values relative to both axes, the green dashed line $f_{\rm max}$, and the blue dash-dotted line $f_{\rm min}$.}
     \label{fig:Upsilon_f1f3_vs_f2}
 \end{figure}

We numerically evaluate the three-point autocorrelation function, $\Upsilon(f_1, f_2, f_3)$, defined by (\ref{finangc}-\ref{eq:beta_def}) to look into some of its properties. For context in the main text, the properties of $\Upsilon(f_1, f_2, f_3)$ can also be viewed according to its action as a kernel in a three-dimensional integral:
\begin{equation}
    \int_{a_1}^{a_2} df_1 \int_{b_1}^{b_2} df_2 \int_{c_1}^{c_2} df_3 \, {\cal G}(f_1, f_2, f_3) \Upsilon(f_1, f_2, f_3) \,,
\end{equation}
where ${\cal G}$ is an arbitrary function, and the constants $a_i, b_i, c_i \geq 0$ determine the volume $| (a_2-a_1) \times (b_2 - b_1) \times (c_2 - c_1) |$ cuboid region where the integral is evaluated.

First off, it is clear that the function $\Upsilon(f_1, f_2, f_3)$ is symmetric with respect to the interchange of its first two arguments, or that it respects the symmetry $f_1 \leftrightarrow f_2$. Following this, we can look into the properties of $\Upsilon(f_1, f_2, f_3)$ by focusing on variations across the first and third axes, or $f_1$ and $f_3$, with fixed $f_2$.

It is very useful to remark that $f_1 f_2 f_3 \Upsilon(f_1, f_2, f_3)$ is a function of a single variable $\Delta \equiv \Delta(f_1, f_2, f_3) = (f_1^2 + f_2^2 - f_3^2)/(2 f_1 f_2)$. This dependence is easier to understand. Furthermore, because of the factor $\beta(\Delta)$ defined as \eqref{eq:beta_def} (a top-hat function that accepts only contributions in $\Delta \in [-1, 1]$), we only need to evaluate $f_1 f_2 f_3 \Upsilon(f_1, f_2, f_3)$ for $\Delta \in [-1, 1]$. For this purpose, we define the normalized three-point autocorrelation function, $X(\Delta)$, as
\begin{equation}
    X(\Delta) \equiv X\left(\Delta(f_1, f_2, f_3)\right) = (\pi/2) f_1 f_2 f_3 \Upsilon(f_1, f_2, f_3) \,.
\end{equation}
This is shown in Figure \ref{fig:Upsilon_domain}. Obviously, $(\pi/2) f_1 f_2 f_3 \Upsilon(f_1, f_2, f_3) \sim {\cal O}(1)$ when $\Delta(f_1, f_2, f_3) \in [-1, 1]$, and vanishes otherwise. Generally the shape of $X(\Delta)$ is of a quartic polynomial because the Legendre polynomial of highest order that enters the expression is $P_4(x)$. It is easy to show that
\begin{equation}
\label{eq:normalized_3p_acf}
    X(\Delta) =
    \begin{cases}
        \dfrac{2 \pi^2}{315} \left( 3 + \Delta^{2} \right)^2 & \,\,\,\, , \,\,\,\, |\Delta| < 1 \phantom{\dfrac{1}{1}} \\
        \dfrac{16 \pi^2}{315} & \,\,\,\, , \,\,\,\, \Delta = \pm 1 \phantom{\frac{\dfrac{1}{1}}{\dfrac{1}{1}}}\\
        0 & \,\,\,\, , \,\,\,\, |\Delta| > 1 \phantom{\dfrac{1}{1}} \,.
    \end{cases}
\end{equation}
Moreover, we can see that $(\pi/2) f_1 f_2 f_3 \Upsilon(f_1, f_2, f_3)$, although piecewise in $\Delta \in (-\infty, \infty)$, is symmetric with respect to $\Delta$, or that $X(\Delta) = X(-\Delta)$; this can be inferred using (\ref{bessel}-\ref{eq:beta_def}), or by referring to the parity properties of the Legendre polynomials. For $\Delta \in (-1, 1)$, the minimum of $X(\Delta)$ of $ 2\pi^2/35 \sim 0.564$ occurs at $\Delta=0$, or in terms of the frequencies, when $f_1^2 + f_2^2 - f_3^2 = 0$, and the maximum of $32 \pi^2/315 \sim 1.003$ at $\Delta=1^-$ or when $f_1^2 + f_2^2 - f_3^2 = 2 f_1 f_2$. The function values of the quartic polynomial $X(\Delta)$ transition smoothly between this minimum and maximum in $\Delta \in (-1, 1)$. The entire range and domain of the three-point autocorrelation function, $\Upsilon(f_1, f_2, f_3)$, can be deduced based on the above properties and Figure \ref{fig:Upsilon_domain}. 

Algorithmically, $\Upsilon(f_1, f_2, f_3)$ can be computed as follows: map given points $(f_1, f_2, f_3) \in \mathbb{R}^3$ to $\Delta \in \mathbb{R}$ using $\Delta \equiv \Delta(f_1, f_2, f_3) = (f_1^2 + f_2^2 - f_3^2)/(2 f_1 f_2)$, find the value of $X$ corresponding to the computed $\Delta$'s using \eqref{eq:normalized_3p_acf}, and compute $\Upsilon(f_1, f_2, f_3) = (2/\pi) X\left( \Delta(f_1, f_2, f_3) \right) / (f_1 f_2 f_3)$. Because the polynomial $X(\Delta)$ is dimensionless, clearly the units of the three-point autocorrelation function is given by the inverse of the volume $f_1 f_2 f_3$ in frequency space. For nanohertz frequencies, $f \sim 0.1 \, {\rm yr}^{-1} \sim 3.2 \times 10^{-9} \, {\rm Hz}$, this gives $\Upsilon \sim 10^{26} \, {\rm s}^3$. The following numerical evaluations will confirm this estimated order of magnitude, relevant for PTA science. Figure \ref{fig:Upsilon_1d} shows the diagonals, $f_1=f_2=f_3$, sub-diagonals $f_1 = f_2$ and $f_2=f_3$, and off-diagonal frequency components of $\Upsilon(f_1, f_2, f_3)$ at nanohertz frequencies.

We can see that the diagonals, $f_1=f_2=f_3$,  contribute as a power law that increases indefinitely at low frequencies. This contribution to the integral is regulated by cutting of the power below some threshold frequency. PTA analysis do this by projecting out long time deterministic trends in the data, effectively cutting off power at $f \lesssim 1/T$ by  a transmission function that factors as $\sim f^6$. This suppresses contributions to the total background power of GWs with frequencies below a nanohertz, or periods longer than three decades. We can also see that the sub-diagonal contributions, $\Upsilon(f, f, f_{\rm ref})$ and $\Upsilon(f_{\rm ref}, f, f)$, act as power laws. However, compared with the main diagonals, we can see that the sub-diagonals are automatically cut-off at a frequency roughly half the reference frequency. Then, for typical SGWB sources in PTA that have a red spectra, the sub-diagonal components can be expected to give a convergent contribution to the total integral.

Figure \ref{fig:Upsilon_1d} also shows off-diagonal components, $\Upsilon(f, f_{\rm ref}, f_{\rm ref})$ and $\Upsilon(f_{\rm ref}, f_{\rm ref}, f)$. These components factor in as a power law that scales as $\sim f^{-1}$ that indefinitely extends to lower frequencies, but with a cut-off at roughly twice the reference frequency. It is useful to note that $f \Upsilon(f, f_{\rm ref}, f_{\rm ref}) \sim$Constant and $f \Upsilon(f_{\rm ref}, f_{\rm ref}, f) \sim$Constant at $f \lesssim f_{\rm ref}/2$, and so these off-diagonal components also add in to a convergent contribution to the total integral.

Figures \ref{fig:Upsilon_f1f2_vs_f3}-\ref{fig:Upsilon_f1f3_vs_f2} highlight further off-diagonal components of $\Upsilon(f_1, f_2, f_3)$ that are away compared to the main and sub-diagonals. The dark shaded regions in \ref{fig:Upsilon_f1f2_vs_f3}-\ref{fig:Upsilon_f1f3_vs_f2} are supportive of the analytic properties that the three-point autocorrelation function reaches its relevant values with $\Delta = (f_1^2 + f_2^2 - f_3^2)/(2f_1f_2) \leq 1$ and vanishes elsewhere.

\bibliography{refs}

\end{document}